\documentclass[nohyper,notoc]{JHEP}

\usepackage{amsmath,euscript,array,amssymb,cite} 
\def\bd{\begin{document}}
\def\ed{\end{document}}
\let\bm=\bibitem
\let\la=\label
\newcommand{\mxcom}[2]{[ #1 , #2 \}}
\newcommand{\acom}[2]{\{ #1 , #2 \}}
\newcommand{\com}[2]{[ #1 , #2 ]}
\newcommand{\dspst}[1]{\displaystyle #1}
\newcommand{\scst}[1]{{\scriptstyle #1}}
\newcommand{\sscst}[1]{{\scriptscriptstyle #1}}
\newcommand{\rf}[1]{(\ref{#1})}
\newcommand{\hoch}[1]{$\, ^{#1}$}

\newcommand{\EQ}[1]{\begin{equation} #1 \end{equation}}
\newcommand{\AL}[1]{\begin{subequations}\begin{align} #1 
\end{align}\end{subequations}}
\newcommand{\SP}[1]{\begin{equation}\begin{split} #1 \end{split}\end{equation}}
\newcommand{\ALAT}[2]{\begin{subequations}\begin{alignat}{#1} #2 
\end{alignat}\end{subequations}}
\newcommand{\coset}[2]{\frac{#1}{#2}}

\newcounter{multieqs}
\newcommand{\eqnumber}{\addtocounter{multieqs}{1}
\addtocounter{equation}{-1}} \newcommand{\begalph}
{\setcounter{multieqs}{0} \addtocounter{equation}{1}
\renewcommand{\theequation}
{\thesection.\arabic{equation}.\alph{multieqs}}}
\newcommand{\alphend} {\setcounter{multieqs}{0}
\renewcommand{\theequation}{\thesection.\arabic{equation}}}

\def\be{\begin{equation}}
\def\ee{\end{equation}}
\newcommand{\eq}[1]{(\ref{#1})}
\def\nn{\nonumber}
\def\bea{\begin{eqnarray}}
\def\eea{\end{eqnarray}}

\def\bq{\begin{equation}}
\def\fq{\end{equation}}
\def\bqr{\begin{eqnarray}}
\def\fqr{\end{eqnarray}}
\def\non{\nonumber \\}
\def\noi{\noindent}
\newcommand{\lab}[1]{\label{#1}}

\def\npb#1#2#3{Nucl. Phys. {\bf{B#1}} #3 (#2)}
\def\plb#1#2#3{Phys. Lett. {\bf{#1B}} #3 (#2)}
\def\prl#1#2#3{Phys. Rev. Lett. {\bf{#1}} #3 (#2)}
\def\prd#1#2#3{Phys. Rev. {D \bf{#1}} #3 (#2)}
\def\cmp#1#2#3{Comm. Math. Phys. {\bf{#1}} #3 (#2)}
\def\cqg#1#2#3{Class. Quantum Grav. {\bf{#1}} #3 (#2)}
\def\nppsa#1#2#3{Nucl. Phys. B (Proc. Suppl.) {\bf{#1A}}#3 (#2)}
\def\ap#1#2#3{Ann. of Phys. {\bf{#1}} #3 (#2)}
\def\ijmp#1#2#3{Int. J. Mod. Phys. {\bf{A#1}} #3 (#2)}
\def\rmp#1#2#3{Rev. Mod. Phys. {\bf{#1}} #3 (#2)}
\def\mpla#1#2#3{Mod. Phys. Lett. {\bf A#1} #3 (#2)}
\def\jhep#1#2#3{J. High Energy Phys. {\bf #1} #3 (#2)}
\def\atmp#1#2#3{Adv. Theor. Math. Phys. {\bf #1} #3 (#2)}

\newcommand{\PD}[2]{\frac{\partial #1}{\partial #2}}
\newcommand{\MAT}[1]{\begin{pmatrix}#1\end{pmatrix}}
\def\pa{\partial} \def\del{\partial}
\def\delb{\bar{\partial}}
\def\xx{\times}
\def\trp{^{\top}}
\def\inv{^{-1}}
\def\dag{^{\dagger}}
\def\pr{^{\prime}}
\def\rar{\rightarrow}
\def\lar{\leftarrow}
\def\lrar{\leftrightarrow}
\def\hlf{\frac{1}{2}}
\def\ove#1{\frac{1}{#1}}
\def\Box{\square}
\def\two{{(2)}}
\def\eighth{\tfrac18}
\def\fourth{\tfrac14}
\def\hf{\tfrac12}
\def\sqrtwo{\sqrt2}
\def\ft#1#2{{\textstyle{{\scriptstyle #1}\over {\scriptstyle #2}}}}
\newcommand{\ex}[1]{{\rm e}^{#1}}

\def\tr{{\rm tr}}
\def\trN{{\rm Tr}_N} \def\TrN{{\rm Tr}_N}
\def\trtwo{\tr^{}_2\,}
\def\aD{{\dot\alpha}} \def\dalpha{{\dot\alpha}}
\def\bD{{\dot\beta}}  \def\dbeta{{\dot\beta}}
\def\sD{{\dot\sigma}}  \def\gD{{\dot\gamma}}
\def\dD{{\dot\delta}}

\def\ddt{\dot\delta}

\def\a{\alpha}      \def\da{{\dot\alpha}}
\def\b{\beta}       \def\db{{\dot\beta}}
\def\c{\gamma}  \def\C{\Gamma}  \def\cdt{\dot\gamma}
\def\d{\delta}  \def\D{\Delta}  \def\ddt{\dot\delta}
\def\e{\epsilon}        \def\vare{\varepsilon}
\def\f{\phi}    \def\F{\Phi}    \def\vvf{\f}
\def\h{\eta}
\def\k{\kappa}
\def\l{\lambda} \def\L{\Lambda}
\def\m{\mu} \def\n{\nu}
\def\p{\pi} 
\def\r{\rho}
\def\s{\sigma}  
\def\t{\tau}
\def\th{\theta} \def\Th{\Theta} \def\vth{\vartheta}
\def\X{\Xeta}
\def\z{\zeta}

\def\psibar{\bar\psi}
\def\Nbar{\bar{\cal N}}
\def\Ubar{\bar U}
\def\Mbar{\bar{\cal M}}
\def\Deltabar{\bar\Delta}
\def\wbar{\bar w}
\def\mubar{\bar\mu}
\def\abar{\bar a}
\def\bbar{\bar b}
\def\sigmabar{\bar\sigma}
\def\etabar{\bar\eta}
\def\zetabar{\bar\zeta}
\def\mubar{\bar\mu}
\def\nubar{\bar\nu}
\def\zb{\bar{z}}     \def\Zb{\bar{Z}}
\def\bz{\bar{z}}

\def\cA{{\cal A}} \def\cB{{\cal B}} \def\cC{{\cal C}}
\def\cD{{\cal D}} \def\cE{{\cal E}} \def\cF{{\cal F}}
\def\cG{{\cal G}} \def\cH{{\cal H}} \def\cI{{\cal I}}
\def\cJ{{\cal J}} \def\cK{{\cal K}} \def\cL{{\cal L}}
\def\cM{{\cal M}} \def\cN{{\cal N}} \def\cO{{\cal O}}
\def\cP{{\cal P}} \def\cQ{{\cal Q}} \def\cR{{\cal R}}
\def\cS{{\cal S}} \def\cT{{\cal T}} \def\cU{{\cal U}}
\def\cV{{\cal V}} \def\cW{{\cal W}} \def\cX{{\cal X}}
\def\cY{{\cal Y}} \def\cZ{{\cal Z}}

\def\A{{\cal A}}
\def\J{{\cal J}}
\def\D{{\cal D}}
\def\G{{\cal G}}
\def\K{{\cal K}}
\def\CM{{\cal M}} \def\CN{{\cal N}}
\def\M{{{\cal M}'}}
\def\N{{\cal N}} \def\P{{\cal P}}
\def\Q{{\cal Q}}

\def\ii{{\rm i}} \def\jj{{\rm j}} \def\kk{{\rm k}} \def\llr{{\rm l}}

\def\ua{\underline{\alpha}}
\def\ub{\underline{\phantom{\alpha}}\!\!\!\beta}
\def\uc{\underline{\phantom{\alpha}}\!\!\!\gamma}
\def\um{\underline{\mu}}
\def\ud{\underline\delta}
\def\ue{\underline\epsilon}
\def\una{\underline a}\def\unA{\underline A}
\def\unb{\underline b}\def\unB{\underline B}
\def\unc{\underline c}\def\unC{\underline C}
\def\und{\underline d}\def\unD{\underline D}
\def\une{\underline e}\def\unE{\underline E}
\def\unf{\underline{\phantom{e}}\!\!\!\! f}\def\unF{\underline F}
\def\unm{\underline m}\def\unM{\underline M}
\def\unn{\underline n}\def\unN{\underline N}
\def\unp{\underline{\phantom{a}}\!\!\! p}\def\unP{\underline P}
\def\unq{\underline{\phantom{a}}\!\!\! q}
\def\unQ{\underline{\phantom{A}}\!\!\!\! Q}
\def\unH{\underline{H}}

\def\As {{A \hspace{-6.4pt} \slash}\;}
\def\Ds {{D \hspace{-6.4pt} \slash}\;}
\def\ds {{\del \hspace{-6.4pt} \slash}\;}
\def\ss {{\s \hspace{-6.4pt} \slash}\;}
\def\ks {{ k \hspace{-6.4pt} \slash}\;}
\def\ps {{p \hspace{-6.4pt} \slash}\;}
\def\pas {{{p_1} \hspace{-6.4pt} \slash}\;}
\def\pbs {{{p_2} \hspace{-6.4pt} \slash}\;}

\def\Dbarslash{\,\,{\raise.15ex\hbox{/}\mkern-12mu {\bar\D}}}
\def\Dslash{\,\,{\raise.15ex\hbox{/}\mkern-12mu \D}}
\def\delslash{\,\,{\raise.15ex\hbox{/}\mkern-9mu \partial}}
\def\delbarslash{\,\,{\raise.15ex\hbox{/}\mkern-9mu {\bar\partial}}}

\def\Fh{\hat{F}}
\def\Xh{\hat{X}}
\def\ah{\hat{a}}
\def\xh{\hat{x}}
\def\yh{\hat{y}}
\def\ph{\hat{p}}
\def\xih{\hat{\xi}}
\def\bfhat#1{{\bf \hat{#1}}}

\def\pt{\tilde{p}}
\def\Psit{\tilde{\Psi}}

\newcommand{\BN}{\boldsymbol{N}}
\newcommand{\BR}{\boldsymbol{R}}
\newcommand{\Bk}{\boldsymbol{k}}
\newcommand{\BL}{\boldsymbol{L}}
\newcommand{\BH}{\boldsymbol{H}}
\newcommand{\Bone}{\boldsymbol{1}}
\newcommand{\Be}{\boldsymbol{e}}
\newcommand{\Balpha}{{\boldsymbol{\alpha}}}
\newcommand{\Bxi}{{\boldsymbol{\xi}}}
\newcommand{\Bomega}{{\boldsymbol{\omega}}}
\newcommand{\BI}{\boldsymbol{I}}
\newcommand{\BJ}{\boldsymbol{J}}
\newcommand{\BK}{\boldsymbol{K}}

\def\VEV#1{\left\langle #1\right\rangle}
\def\braket#1#2{\VEV{#1 | #2}}
\def\VEV#1{\left\langle #1\right\rangle}
\def\Vev#1{\big\langle{#1}\big\rangle}
\let\vev=\Vev
\def\grp{{\EuScript U}}
\newcommand{\bra}[1]{\langle #1|}
\newcommand{\ket}[1]{|#1 \rangle}
\newcommand{\ipr}[2]{\langle #1 | #2 \rangle}
\newcommand{\xpv}[1]{\langle #1  \rangle}
\def\vac{|0\rangle}
\def\lvac{\langle 0|}

\def\det{{\rm det}}
\def\SU{\text{SU}}
\def\U{\text{U}}
\def\HK{hyper-K\"ahler}

\def\SU{\text{SU}}
\def\SO{\text{SO}}
\def\O{\text{O}}
\def\Gl{\text{Gl}}
\def\Sp{\text{Sp}}
\def\Tr{\text{Tr}}
\def\USp{\text{USp}}

\def\sst{\scriptscriptstyle}
\newcommand{\BOX}[1]{\fbox{\parbox{15cm}{#1}}}
\def\zo{{\sst(0)}}
\def\Pinfty{{\cal P}_\infty}
\newcommand{\0}{\,\!}      
\def\one{1\!\!1\,\,}
\def\im{\imath}
\def\jm{\jmath}

\newcommand{\slsh}[1]{/ \!\!\!\! #1}

\def\One{{1}}
\def\Weff{{\cal W}_{\rm effective}}
\def\ms{{\mathfrak M}}
\def\ns{{\mathfrak N}}
\def\cms{\widehat{\mathfrak M}}
\def\F{{\mathfrak F}}
\def\Q{{\mathfrak Q}}
\def\Z{{\EuScript Z}}
\def\CC{{\EuScript C}}
\def\g{{\EuScript g}}
\def\I{{\EuScript I}}
\def\FF{F}
\def\B{{\EuScript B}}
\def\NS{{\EuScript N}}
\def\AA{A}
\def\ZZ{\mathbb{Z}}
\def\bcomment#1{}

\title{Notes on Noncommutative Instantons}
\author{Chong-Sun Chu$^{a}$, Valentin V. Khoze$^b$
and Gabriele Travaglini$^b$\\
$^a$Centre for Particle Theory, Department of Mathematical Sciences,\\
University of Durham, Durham, DH1 3LE, UK\\
$^b$Centre for Particle Theory
and Institute for Particle Physics Phenomenology,\\
Department of Physics, University of Durham,
Durham, DH1 3LE, UK \\
E-mail: {\tt chong-sun.chu, valya.khoze,
gabriele.travaglini@durham.ac.uk }}
 
\abstract{We study in detail the ADHM construction of $U(N)$ instantons on 
noncommutative Euclidean space-time $\bf{R}_{\rm NC}^4$ and noncommutative space
$\bf{R}_{\rm NC}^2\times \bf{R}^2$. We point out that the completeness condition
in the ADHM construction could be invalidated in certain circumstances. 
When this happens, regular instanton configuration may not exist even if the
ADHM constraints are satisfied. Some of the existing solutions in the literature
indeed violate the completeness condition and hence are not correct. 
We present alternative  solutions for these cases. In particular, we show
for the first time how to construct explicitly regular $U(N)$ instanton
solutions on $\bf{R}_{\rm NC}^4$ and on $\bf{R}_{\rm NC}^2\times \bf{R}^2$.
We also give a simple general argument based on the Corrigan's identity that
the topological charge of noncommutative regular instantons  
is always an integer. 
}

\keywords{Noncommutative Geometry, Solitons Monopoles and Instantons}

\preprint{{\tt hep-th/0108007}}

\begin{document}

\section{Introduction}\label{sec:S1}

There has been a lot of interest in gauge theories on
noncommutative spaces. One of the reasons for this interest is the natural
appearance of noncommutativity $[x^\mu, x^\nu]=i\theta^{\mu\nu}$
in the framework of
string theory and D-branes \cite{DHull,Chu:1999qz,Schomerus:1999ug,SWnc}.
Noncommutative gauge theories are also fascinating  on their own
right \cite{DNrev}, 
mostly due to a mixing between the infrared (IR) and the ultraviolet
(UV) degrees of freedom discovered in \cite{Minwalla}.
While UV/IR
mixing arises at the perturbative level, it has been suggested that
a resolution of it may lie at the nonperturbative level by resumming
nonplanar diagrams at all loops
\cite{Minwalla,fischler,gubser,griguolo}. In gauge theories,
one would also need to take into account the effects of noncommutative
instantons on the IR physics \cite{Armoni:2001br,HKT,Chu:2001fe}. 
Noncommutative instantons play an essential role in the
understanding of the nonperturbative physics of noncommutative gauge
theories. In fact, the considerations in e.g. \cite{HKT}
require the existence of non-singular noncommutative instanton  
configurations.
The objective of this paper is to show that such
instanton solutions can always be constructed for both
space-time and space-space noncommutativity.

Employing the formalism Atiyah, Drinfeld, Hitchin and Manin (ADHM)
\cite{ADHM},
Nekrasov and Schwarz constructed \cite{NS} the first (singular)
instanton on 
noncommutative $\bf{R}^4$. They showed that the ADHM constraints are
modified by the noncommutativity. Solving the ADHM constraints, they
obtained an anti-selfdual configuration whose singularity  
at short distances is regulated by
the noncommutative scale. Following this pioneering work, 
the role of the projectors in the noncommutative ADHM construction 
was then clarified in a series of papers by Furuuchi
\cite{Furuuchi1,Furuuchi2,Furuuchi3}. Other related works appeared  in
\cite{pmho,LTY,KLY,Nekrasov:2000zz,Schwarz:2001ru,Konechny:2001wz,CLMS,
Harvey,K2}.

As will be discussed in  \S\ref{sec:S2}
two important elements in the ADHM construction are the 
{\it factorization} condition  \eq{dbd}
and the {\it completeness} relation \eq{cmpl}.   It is well
known that the  factorization condition  amounts to the famous ADHM
constraints. As it was
shown in the original work of  Nekrasov and Schwarz \cite{NS}, the same
is true in the noncommutative case and the  ADHM constraints are modified
in a relatively simple manner when noncommutativity is turned on. 
Solving the noncommutative ADHM constraints has been the focus in the literature
since the work of \cite{NS}. However, while the completeness relation is always
guaranteed to hold in the commutative case, it is no
longer so when one turns on noncommutativity. The breakdown of this
completeness relation is closely related to the fact that some
states are often projected out in the construction of the
non-singular noncommutative instantons. 

It will turn out that the completeness relation can always be satisfied
for the case of space-time noncommutativity and regular instanton 
solutions can be found.
However, in the earlier version of this paper we 
have met an obstacle in finding non-singular instanton configurations 
in the case of space-space noncommutativity due to the breakdown of
the completeness relation. 
Subsequently, it was proposed in \cite{K2} that non-singular solutions 
with space-space noncommutativity can be constructed in the formalism
where no states are projected out. In this formalism the completeness
relation is always satisfied, 
but the resulting instanton solutions appear to be singular.
Motivated by the fact that the singularities disappeared
in certain gauge-invariant quantities, the authors of \cite{K2}
conjectured that these singularities can always be removed by 
gauge transformations and regular instanton solutions do exist.
In what follows we will prove that this is indeed the case
by constructing these non-singular solutions explicitly.

The paper is organized as follows.
In  \S\ref{sec:S12}
we give a basic review of the properties of noncommutative
space-time. In particular we consider two cases:
(1) the noncommutative  Euclidean space-time   
$\bf{R}_{\rm NC}^4$ with selfdual $\th$ (SD-$\th$), and 
(2) the noncommutative space $\bf{R}_{\rm NC}^2\times \bf{R}^2$. 

In  \S\ref{sec:S2} we review and discuss the ADHM construction
of noncommutative instantons. In  \S\ref{sec:S22} 
we focus on discussing the factorization condition  
and the completeness relation.  
We show that the  factorization condition  amounts to the modified ADHM
constraints and 
we find that solutions of the ADHM constraints do not necessarily guarantee
the completeness relation. It  is an independent condition and we give a
necessary and sufficient condition \eq{reqt}
for the completeness relation to be valid. In  \S\ref{sec:S23},
we use the Corrigan's identity to
give a general argument that the topological charge of a
noncommutative instanton is always an integer, $|Q|=k.$

We then turn our attention to  explicit examples to illustrate these points.
In  \S 4 
the construction of single $U(1)$ instanton configurations on
$\bf{R}_{\rm NC}^4$ and $\bf{R}_{\rm NC}^2\times \bf{R}^2$ is  studied.
We show that  anti-selfdual (ASD) 
ADHM instantons exist on $\bf{R}_{\rm NC}^4$ with SD-$\th$,
but not the SD-instantons. Explicit non-singular ASD instantons
are found using the formalism where certain states from the Hilbert space
are projected out.
For the case of $\bf{R}_{\rm NC}^2\times \bf{R}^2$, we 
find that
the completeness relation is violated in this formalism.
We then modify our approach and describe a construction on 
$\bf{R}_{\rm NC}^2\times \bf{R}^2$
where no states are projected out, the completeness relation holds and
regular instanton solutions can be found explicitly.
The same is true for higher $U(N)$.
In  \S 5 we study the ASD $U(1)$ two-instanton solution. 
The topological charge is computed directly and is
found to be an integer, $Q=-2.$ 
Our result is different from a previous analysis
\cite{KLY}--versions 1-3,
which reported a non-integral topological charge. 

In  \S 6 we consider a general $U(N)$ gauge group and construct
 ASD and SD instantons on SD-$\bf{R}_{\rm NC}^4$ and instantons
on $\bf{R}_{\rm NC}^2 \times \bf{R}^2$.
For
the case of ASD instanton, we find that the existing ansatz in
the literature does not satisfy the completeness relation. Quite
amazingly, an alternative ansatz \eq{Ucor} can be found, which
does not contain an overall factor of the shift operator $u^\dagger$
as in the original ansatz 
\eq{udefff}, but satisfies the completeness relation and leads to a regular
instanton solution.  
The topological charge of the ASD 1-instanton solution is explicitly computed 
and found to be equal to minus one.
For the case of SD-instanton, the solution is regular and
there is no need to introduce any projector. 
We also note that at large distances (or in the small noncommutativity limit)  
the SD/SD instanton approaches the regular-gauge BPST instanton, 
while the ASD/SD
instanton tends to the  singular-gauge BPST
anti-instanton.  

\section{Noncommutative $\bf{R}^4$}\label{sec:S12}

We will work in flat Euclidean space-time $\bf{R}^4$ with
noncommutative coordinates $x^m$ which satisfy the commutation
relations \EQ{ [x^m, x^n] = i \th^{mn}\ , \label{crxdef}} where
$m,n=1,2,3,4$ are the Euclidean Lorentz indices and $\th^{mn}$ is
an antisymmetric real constant matrix. Using Euclidean space-time
rotations, $\th^{mn}$ can be always brought to the form \EQ{
\th^{mn}=
\begin{pmatrix} 0 & \th^{12} & 0 & 0 \\
-\th^{12}& 0 & 0 & 0 \\ 0 & 0 & 0 &\th^{34}\\
 0 & 0 & -\th^{34}& 0 \end{pmatrix} \ .
\label{thgen}}
In terms of complex coordinates
\SP{
z_1 &= x_2+ i x_1 \ , \quad  \zb_1 = x_2- i x_1 \ , \\
z_2 &=  x_4+ ix_3 \ , \quad  \zb_2 = x_4- ix_3 \ ,
\label{zzbdef}}
the commutation relations \eqref{crxdef} take the form
\SP{
[z_1,\zb_1] &= - 2\th^{12} \ , \quad [z_i,z_j] = 0 \ , \\
[z_2,\zb_2] &= - 2\th^{34}  \ , \quad [z_i,\zb_{j\neq i}]=0 \ ,
\label{crss}}
where $i,j =1,2$  denote the indices for the complex coordinates.

There are three important cases to consider:
\begin{itemize}
\item When $\th^{12}=0=\th^{34}$ all the commutators vanish giving the
ordinary commutative $\bf{R}^4$. The corresponding world-volume 
gauge theory is
the commutative gauge theory, and instanton solutions are given
by the standard ADHM construction \cite{ADHM,CFGT,CWS,Corrigan:1979di,OSB}.
Recent reviews and
applications of the ADHM calculus can be found in \cite{KMS,DHKMV}.
\item 
When either $\th^{12}$ or $\th^{34}$ vanishes, 
the matrix $\th^{mn}$ is of rank-two. This case corresponds to
the direct product of the ordinary commutative 2-dimensional space with
the noncommutative 2-dimensional space,
$\bf{R}_{\rm NC}^2\times \bf{R}^2$. For definiteness we set here $\th^{34}=0$
and introduce the notation $\th^{12}\equiv -\zeta/2$ in such a way  that
\EQ{[z_1,\zb_1] = - \zeta \ , \quad [z_2,\zb_2] = 0\ ,
\quad [z_i,z_j] = 0 \ .
\label{crnc2}}
Physical applications
of this situation  involve world-volume gauge theories with noncommutative
space and commutative time.\footnote{
It easily follows from \eqref{thgen} that the {\it general} description of
noncommutative 3-dimensional space
and commutative time is given by $\bf{R}_{\rm NC}^2\times\bf{R}^2$.}
\item A rank-four matrix $\th^{mn}$ (with $\th^{12}\neq 0$ and
$\th^{34}\neq 0$) generates the noncommutative Euclidean space-time
$\bf{R}_{\rm NC}^4=\bf{R}_{\rm NC}^2\times \bf{R}_{\rm NC}^2$.
The corresponding world-volume gauge theory has
noncommutative (Euclidean) time. Since both components of $\theta$ are
non-vanishing, they can be made equal, $\th^{12}=\th^{34}\equiv -\zeta/4,$
with appropriate rescalings of the four
coordinates $x^m$ and, if necessary, a parity transformation.\footnote{
Physics of noncommutative space-time is determined by $\th^{mn}$ and
certainly is not invariant under dilatations and parity transformations.
However the general case can be always recovered from the simple case
$\th^{12}=\th^{34}$ via opposite rescalings.}
Equations \eqref{crss} become
\EQ{
[z_i,\zb_j] = - \frac{\zeta}{2} \d_{ij} \ , \quad [z_i,z_j] = 0\ .
\label{crsd}}
In fact, the condition $\th^{12}=\th^{34}$ gives
a selfdual (SD) theta, ${\hf}\epsilon^{mnkl}\th_{kl}=\th^{mn}$, while
the anti-selfdual (ASD) theta, ${\hf}\epsilon^{mnkl}\th_{kl}=-\th^{mn}$,
corresponds to $\th^{12}=- \th^{34}$. From now on when discussing the
$\bf{R}_{\rm NC}^4$ case we will assume the space-time rescaling leading to the
SD-theta, \eqref{crsd}.\footnote{The ASD-theta can be obtained from this by
a parity transformation of two coordinates. This would also change the sign of
the selfduality of the instanton configurations in the world-volume gauge
theory.}
\end{itemize}
The commutation relations \eqref{crnc2} and \eqref{crsd} imply that the
space-time coordinates $x^m$ should be thought of as operators
acting on a Hilbert space. The operator language and its
Hilbert space representation will be discussed below in 
\S\ref{sec:S13}.
In the following sections we will be concerned with constructing instanton
configurations as the solutions to the operator-valued
(anti)-selfduality equations \eqref{asdeq}.
Following Nekrasov, Schwarz and Furuuchi,
\cite{NS,Furuuchi1,Furuuchi2,Furuuchi3} we will use the
operator-valued ADHM construction to determine the solutions of \eqref{asdeq}.
In items 2 and 3 above we chose our notation in such a way that 
$\th^{12}+\th^{34}=\hf \zeta$,
which will allow us to treat both noncommutative cases simultaneously.
In  \S\ref{sec:S2} and \S\ref{sec:S3} we
will review the ADHM construction of instantons and point out important
subtleties specific to noncommutative cases.

In order to discuss instanton effects in gauge theories on $\bf{R}_{\rm NC}^4$
we require a dictionary between the
operator-valued ADHM instanton expressions
and the ordinary c-number functions which are used for the functional integral
representation of noncommutative gauge theories.
This dictionary is provided by the map between the operators
and the operator symbols outlined in \S\ref{sec:S14}.

In the semiclassical approximation to noncommutative gauge
theories one expands the action around the minima of the Euclidean
action -- the noncommutative c-number-instantons -- and integrates
over the c-number-fluctuations around the instantons. The
single-instanton integration measure can be in principle
determined following the standard commutative instanton analysis
of \cite{tHooft, Bernard} by calculating Jacobians for the change
of the integration variables from the fields to the instanton
collective coordinates. The ADHM measure for $2$ noncommutative
$U(1)$ instantons was derived in \cite{LTY}. The general
all-orders in the instanton number ADHM supersymmetric measure was
written down in Refs.~\cite{DHKpart,HKT}. 

\subsection{Operator language}\label{sec:S13}

A Hilbert space representation for the noncommutative geometry \eqref{crnc2}
or \eqref{crsd} can be easily constructed by using complex variables
\eqref{zzbdef} and realizing
$z$ and $\zb$ as creation and annihilation operators in the Fock space
for simple harmonic oscillators (SHO). The fields in a noncommutative
gauge theory are described by functions
of $z_1,\zb_1,z_2,\zb_2$. In the case of $\bf{R}_{\rm NC}^2\times\bf{R}^2$,
the arguments $z_2$ and $\zb_2$ are ordinary c-number coordinates, while
$z_1$ and $\zb_1$ are the creation and annihilation operators of a single
SHO:
\EQ{
z_1 \ket{n}
= \sqrt{\z}\; \sqrt{n+1} \ket{n}\ , \quad
\zb_1 \ket{n}
= \sqrt{\z}\; \sqrt{n} \ket{n-1} \ .
}

The noncommutative space-time
$\bf{R}_{\rm NC}^4=\bf{R}_{\rm NC}^2\times \bf{R}_{\rm NC}^2$
requires two oscillators. The SHO Fock space
$\cH$ is spanned by the basis $\ket{n_1,n_2}$ with $ n_1,n_2 \geq 0 $:
\bea
&z_1 \ket{n_1,n_2}
= \sqrt{\frac{\z}{2}}\; \sqrt{n_1+1} \ket{n_1+1,n_2}\ , \quad
z_2 \ket{n_1,n_2}
= \sqrt{\frac{\z}{2}}\; \sqrt{n_2+1} \ket{n_1,n_2+1}\ , \nn \\
&\zb_1 \ket{n_1,n_2}
= \sqrt{\frac{\z}{2}}\; \sqrt{n_1} \ket{n_1-1,n_2}\ , \quad
\zb_2 \ket{n_1,n_2}
= \sqrt{\frac{\z}{2}}\; \sqrt{n_2} \ket{n_1,n_2-1}\ .
\eea
Derivatives of a function $f(z_1,\zb_1,z_2,\zb_2)$ are defined by
\EQ{
\del_i f = \frac{2}{\z} [\zb_i,f]\ ,\quad \delb_i f = -\frac{2}{\z} [z_i,f]\ ,
}
and satisfy the standard requirements for $f=z_j$ or $f=\zb_j$, as well as
the chain rule and the useful identity for the derivative of the
inverse function:
\EQ{\del_i f^{-1} = - f^{-1}(\del_i f)f^{-1} \ , \quad
\delb_i f^{-1} = - f^{-1}(\delb_i f)f^{-1} \ .
\label{dfinv}}
It is convenient to introduce differentials $dz_i$ and $d\zb_i$, which
commute with $z,\zb$'s and anticommute with each other, $dz_i
d\zb_j = -  d\zb_j dz_i$. We also introduce the (anti-)holomorphic exterior
derivatives
$\del, \delb$ and the total exterior derivative $d$,
\EQ{
d = \del + \delb \ , \quad
\del= dz_i \del_i\ , \quad \delb = d\zb_i \delb_i \ ,
}
which satisfy
\EQ{
\del^2 = \delb^2 =0\ , \quad \del \delb = - \delb \del\ , \quad d^2 =0\ .
}
Nonholomorphic derivatives with respect to $x^m$ are defined via
\EQ{
dx^m \frac{\del}{\del x^m}=
d = \del + \delb = dz_i \del_i +  d\zb_i \delb_i \ ,
}
and are explicitly given by
\SP{{\del \over \del x^1} &=i(\del_1-\delb_1) \ , \quad
{\del \over \del x^3} =i(\del_2-\delb_2) \ , \\
{\del \over \del x^2}&=\del_1+\delb_1 \ , \quad
{\del \over \del x^4}=\del_2+\delb_2  \ .
}
We also note that $ dz_1 d\zb_1 dz_2 d\zb_2 = -4\; d^4 x$.

In the complex basis the condition of
selfduality can be easily formulated. A real 2-form $F$
\EQ{ \label{criterion}
F= ( a_1 \; d z_1 d \zb_1 - a_2 \; d z_2 d\zb_2 ) + b \; d z_1 d\zb_2 + \bar{b}
\; d z_2 d \zb_1 + c \; dz_1 dz_2 + \bar{c} \;  d \zb_2 d \zb_1
}
is anti-selfdual iff $a_1 = a_2$ and $c =\bar{c} =0$. ($F$ is
selfdual iff 
$a_1= - a_2$ and $b = \bar{b} =0$.)

The integral on $\bf{R}_{\rm NC}^4$ is defined by the operator trace,
\EQ{
\int d^4 x = (2\pi)^2 \sqrt{\det \th} \; \mbox{Tr} = (\frac{\z \pi}{2})^2 \;
\mbox{Tr} \ . }
In the ordinary commutative case the integral of the total derivative of a
regular function vanishes when the function falls off fast enough at infinity.
In the noncommutative case the integral of a total derivative is the trace
of a commutator. It vanishes only when the trace of each term in the
commutator is individually well-defined (finite). Hence, similarly to the
commutative case, we see that $\int d^4 x \ \del_m K^m$ 
can receive non-vanishing
contributions from the `boundary of integration'.

\subsection{Operator symbols}\label{sec:S14}

Operator symbols are ordinary c-number functions which provide an
alternative to the operator language. Using operator symbols,  the
fields of a noncommutative gauge theory are viewed as ordinary
c-number functions which are multiplied using the star-product:
\EQ{(\Phi * \Psi) (x) \equiv \Phi(x) e^{{i\over 2}\theta^{mn}
\stackrel{\leftarrow}{\partial_m}
\stackrel{\rightarrow}{\partial_n}}  \Psi(x) \ , \label{stardef}}
where here 
$\partial_m$ denotes $\partial/\partial x^m.$ This is
achieved for each of the $\bf{R}_{\rm NC}^2$ factors in 
$\bf{R}_{\rm NC}^2\times \bf{R}^2$ or $\bf{R}_{\rm NC}^4={\rm
NC}\bf{R}^2\times \bf{R}_{\rm NC}^2$ 
via a one-to-one map $\Omega$
from the operators on $\bf{R}_{\rm NC}^2$ to the c-number operator
symbols on $\bf{R}^2.$ This map can be defined \cite{Furuuchi2}
via an inverse normal ordered Fourier transform, 
\EQ{\Omega\ :\
\hat\phi(\hat{x})\ \rightarrow\ \Phi(x)\equiv \int {d^2k \over
(2\pi)^2} e^{ikx} \int d^2 \hat{x} \, :e^{-ik\hat{x}}:\,
\hat\phi(\hat{x})\ , \label{os2d}} 
where hats denote operator
quantities, $\int d^2 \hat{x}$ is the normalized operator trace
$2\pi |\th|\; \mbox{Tr},$ and $:e^{ik\hat{x}}:$ stands for the
normal ordered exponent. This expression for the operator symbol
is particularly transparent in the coherent state basis:
\EQ{\Omega\ :\ \hat\phi(\hat{z},\hat{\zb})\ \rightarrow\
\Phi(z,\zb)= \langle z|\hat\phi(\hat{z},\hat{\zb})|\zb \rangle \ ,
\label{oscs}} 
where $|\zb \rangle$ and $\langle z|$ denote
normalized coherent states, 
\EQ{ \hat{\zb}|\zb \rangle= \zb |\zb
\rangle\ , \quad \langle z| \hat{z} = \langle z| z\ , \quad
\langle z|\zb \rangle=1 \ , } 
and 
\EQ{\langle z|:e^{ik\hat{x}}:|\zb \rangle=e^{ikx} \ . } 
Useful examples of
this correspondence include the expressions for
the operator symbols of the Fock states projectors, 
\EQ{\Omega\ :\
|m\rangle \langle n|\ \rightarrow\ {1\over \sqrt{m!}}\left(z\over
\sqrt{2 |\th|}\right)^m {1\over \sqrt{n!}}\left(\bz\over \sqrt{2
|\th|}\right)^n \ e^{- 2 z\zb/ |\th|} \ , \label{fsos} } 
and the coherent states projectors, 
\EQ{\Omega\ :\ |\bar{w}\rangle \langle
w|\ \rightarrow\ e^{-2(z-w)(\zb-\bar{w})/ |\th|} \ .
\label{csos} }

Using the dictionary above outlined,  
we can easily turn operator-valued expressions
into ordinary functions. As already mentioned,
this is an important requirement for the functional integral
representation of noncommutative gauge theories, 
which uses ordinary functions
(albeit multiplied using the star-product).
The map
between the operators and the operator symbols provides the link between
the ADHM instantons and their semiclassical contributions.

\section{ADHM construction of instantons}\label{sec:S2}

In this section we describe the construction of instantons due to
Atiyah, Drinfeld, Hitchin and Manin (ADHM) \cite{ADHM}, which was
first applied to
noncommutative gauge theories by Nekrasov and Schwarz \cite{NS}.
The commutative ADHM construction was also discussed in 
Refs.~\cite{CFGT,CWS,Corrigan:1979di,OSB,Amati,MO1,MO2,KMS,DHKMV}. 
Here we follow the $U(N)$ formalism of Refs.~\cite{KMS,DHKMV}.

Consider a pure $U(N)$ gauge theory formulated on a  generic
noncommutative Euclidean space (it can be $\bf{R}_{\rm NC}^4$ or
$\bf{R}_{\rm NC}^2\times\bf{R}^2$). The action in the operator language is
given by
\EQ{
S[A]=-\frac1{2g^2}\int d^4x\,{\rm Tr}_N\,F_{mn}F_{mn} \ ,
\label{sec}
}
where $F_{mn}$ is the field strength
$F_{mn}=\partial_mA_n-\partial_nA_m+[A_m,A_n].$
The topological charge $Q$ is defined by
\EQ{
Q=-\frac1{16\pi^2}\int d^4x\,{\rm Tr}_NF_{mn}{}\tilde{F}_{mn} \equiv
-\frac1{16\pi^2}\int d^4x\, \partial_m K^m
\ ,
\label{topc}
}
where $\tilde{F}_{mn}=\tfrac12\epsilon_{mnkl}F_{kl}$
and we used the well-known fact that 
${\rm Tr}_NF_{mn}{}\tilde{F}_{mn}$
can be written as a total derivative. In the ordinary
commutative case the topological charge is an integer equal to
the winding number of the map 
$S^3 \rightarrow S^3$. Here the first $S^3$ is the
boundary at infinity of the space-time $\bf{R}^4$,
and the second $S^3$ is an $SU(2)$ subgroup of the gauge group $U(N)$.
The $k$-instanton configuration is then defined as the general solution of the
(anti)-selfduality equation
\EQ{
F^{mn}=\pm \hf \epsilon^{mnkl}F_{kl} \ 
\label{asdeq}}
in the topological sector $Q=k$. Instantons automatically solve the
nonabelian Maxwell equations thanks to the Jacobi identity and,
hence, give local minima
of the Euclidean action \eqref{sec}:
\EQ{
S_{\rm inst}=\frac{8\pi^2|k|}{g^2} \ .
}
We will see that the topological charge calculated on instanton configurations
in noncommutative $U(N)$ gauge theories is still an integer\footnote{There was
some confusion in the literature concerning this issue. In \cite{Furuuchi2}
Furuuchi argued that the topological charge of noncommutative $U(1)$ instantons
must be integer, however Kim, Lee and Yang in 
\cite{KLY}--versions 1-3
have calculated $Q$ explicitly for the 2-instanton
in $U(1)$ and got a non-integer result. They also obtained non-integer expressions for $Q$ using a $U(2)$ instanton. After the present paper
appeared they have redone their calculations in \cite{KLY}--version 4
obtaining results which now agree with ours.
In addition,
the authors of \cite{CLMS} showed that the noncommutative version of 
't Hooft ansatz, previously introduced in \cite{NS}, does not give
a selfdual configuration, and, hence, its topological charge is not an
integer. An alternative solution presented in \cite{CLMS} does have
an integer $Q$, but at the cost of introducing a non-Hermitian field-strength.
Below we will give a general argument
that for all values of $N$ the topological charge of the noncommutative
$U(N)$ instanton is integer and, moreover, equal to the instanton number $k$.
Then in the following sections we also provide explicit calculations of $Q$
at the 1- and 2-instanton level, finding integer results.
}
for all
values of $N\ge 1$. For $N\ge 2$ one might conjecture that this
result still has topological origins: first,
express $Q$ as an integral of the total derivative $\partial_m K^m$,
and, second, use the c-number operator symbols to evaluate $K^m$.
With an additional assumption that there are no singularities in $K^m$ at
finite values of $x$, one would conclude that $Q$ receives contributions
only from the sphere $S^3$  at spatial infinity,  where
noncommutativity is irrelevant and $Q$ is an integer. This argument, however,
does not explain why $Q$ is an integer for $U(1)$.

When discussing instantons it is very convenient
to introduce a quaternionic notation for the
four-dimensional Euclidean space-time indices
\EQ{
x_{\sst [2] \times [2]} \equiv
x_{\alpha\aD}=x_n\sigma_{n\alpha\aD}\ ,\qquad
\bar x_{\sst [2] \times [2]} \equiv
\bar x^{\aD\alpha}=x_n\bar\sigma_n^{\aD\alpha}\ ,
\label{rae}
}
where $\sigma_{n\alpha\aD}$ are the components of four $2\times2$ matrices
$\sigma_n=(i\vec\tau,1_{\sst[2]\times[2]})$,
and $\tau^c$, $c=1,2,3$ are the three Pauli
matrices. In addition we define the Hermitian conjugate
matrices $\bar\sigma_n=\sigma_n^\dagger=(-i\vec\tau,1_{\sst[2]\times[2]})$
with components $\bar\sigma_n^{\aD\alpha}$.\footnote{
Notice that the spinor indices $\alpha,\aD=1,2$
are raised and lowered with the
$\epsilon$-tensor:
$\bar x^{\aD\alpha}=\epsilon^{\alpha\beta}
\epsilon^{\aD\bD}x_{\beta\bD}$.}
In terms of the complex coordinates \eqref{zzbdef} we have
\EQ{
x_{\alpha\aD}=\begin{pmatrix}z_2 & z_1 \\
-\zb_1& \zb_2 \end{pmatrix}\ ,\qquad
\bar x^{\aD\alpha}=\begin{pmatrix}\zb_2 & -z_1 \\
\zb_1& z_2 \end{pmatrix}\ .
\label{raf}
}
The tensor
\EQ{
\sigma_{mn\,\alpha}{}^{\beta}\equiv \tfrac14(\sigma_{m\alpha\aD}^{}
\sigmabar_n^{\aD\beta}-\sigma^{}_{n\alpha\aD}\sigmabar_m^{\aD\beta})
=\tfrac12 i\eta^a_{mn}\tau^a
\ , \label{sigsd}
}
is selfdual,
$ \tfrac12\epsilon_{mnkl}\sigma_{kl}=\sigma_{mn},$
while
\EQ{
\bar\sigma_{mn}{}^{\aD}_{\bD}\equiv \tfrac14(
\sigmabar_m^{\aD\alpha}\sigma_{n\alpha\bD}^{}
-\sigmabar_n^{\aD\alpha}\sigma_{m\alpha\bD}^{})
=\tfrac12 i\bar\eta^a_{mn}\tau^a
\ , \label{sigasd}
}
is anti-selfdual,
$ \tfrac12\epsilon_{mnkl}\bar\sigma_{kl}=-\bar\sigma_{mn}.$
Here $\eta^a_{mn}$ and $\bar\eta^a_{mn}$ are the standard 't Hooft
$\eta$-symbols.

\subsection{The gauge field and the field strength}\label{sec:S21}

The basic object in the ADHM construction of  selfdual $k$-instantons
(SD-instantons)
is the $(N+2k)\times 2k$ complex-valued matrix
$\Delta_{\sst [N+2k] \times [2k]}$ which is taken to be linear in the
space-time variable $x_n$:\footnote{For
clarity, in this section, we will occasionally show matrix sizes explicitly,
{\it e.g.\/}~the $U(N)$ gauge field will be denoted
$A_{n{\sst [N] \times [N]}}$. To
represent matrix multiplication in this notation
we will underline contracted indices:
$(AB)_{\sst [a] \times [c]} = \ A_{\sst [a] \times \underline{\sst [b]}} \
B_{ \underline{\sst [b]}\times [c]}$. Also we adopt the shorthand
$X_{\,[m}Y_{n]}=X_mY_n-X_nY_m$.}
\begin{equation}
{\rm SD\, instanton:} \quad \Delta_{\sst [N+2k] \times [2k]}(x)\equiv
\Delta_{\sst [N+2k] \times [k] \times [2]} (x)
=a_{\sst [N+2k] \times [k] \times [2]} +
b_{\sst [N+2k] \times [k] \times \underline{\sst [2]}} \
x_{\sst \underline{[2]} \times [2]} \ .
\label{dlt}\end{equation}
Here we have represented the $[2k]$ index set as a product of two index sets
$[k] \times [2]$ and  used a quaternionic representation of
$x_n$ as in \eqref{rae}.
By counting the number
of bosonic zero modes, we will soon verify that $k$ in
Eq.~\eqref{dlt} is indeed the instanton number, while $N$ is the
number of colours of the gauge group $U(N)$.
We further choose
a representation in which $b$ assumes a simple  canonical
form \cite{CFGT}:
\begin{equation}
b_{\sst [N+2k]\times [2k]} =
\begin{pmatrix} 0_{\sst [N]\times [2k]} \\  1_{\sst [2k]\times [2k]}\end{pmatrix}
\ , \qquad
a_{\sst [N+2k]\times [2k]} =
\begin{pmatrix} w_{\sst [N]\times [2k]} \\  a'_{\sst [2k]\times
[2k]}\end{pmatrix}\ .
\label{canform}\end{equation}
As discussed below, the
complex-valued constant matrix $a$ in Eq.~\eqref{dlt}
constitutes a (highly overcomplete) set of collective
coordinates on the instanton moduli space $\ms_k$.

The matrix $\Delta$ used for the construction of the
anti-selfdual instanton (ASD instanton) is given by
\begin{equation}
{\rm ASD\, instanton:} \quad \Delta_{\sst [N+2k] \times [2k]}(x)\equiv
\Delta_{\sst [N+2k] \times [k] \times [2]} (x)
=a_{\sst [N+2k] \times [k] \times [2]} +
b_{\sst [N+2k] \times [k] \times \underline{\sst [2]}} \
\bar x_{\sst \underline{[2]} \times [2]} \ .
\label{dltasd}\end{equation}
It follows from the definitions
that for the SD instanton $\partial_n\Delta = b\sigma_n$,
whereas for the ASD instanton $\partial_n\Delta = b\bar\sigma_n$.

For generic $x$, the null-space of the Hermitian conjugate matrix
$\bar\Delta(x)$
is $N$-dimensional, as it has $N$ fewer rows than columns.
The basis vectors for this null-space can be assembled
into an $(N+2k)\times N$ dimensional  complex-valued matrix
$U(x)$,\footnote{Throughout this, and other sections, an overbar means
hermitian conjugation: $\bar\Delta\equiv\Delta^\dagger$.}
\begin{equation}
\bar\Delta_{\sst [2k] \times \underline{\sst [N+2k]}}
U_{\sst \underline{[N+2k]} \times [N]}
= 0 =
\bar U_{\sst [N] \times \underline{\sst [N+2k]}}
\Delta_{\sst \underline{[N+2k]} \times [2k]} \ ,
\label{uan}\end{equation}
where $U$ is orthonormalized according to
\begin{equation}
\bar U_{\sst [N] \times \underline{\sst [N+2k]}}
 U_{\underline{\sst [N+2k]} \times {\sst [N]}} =  1_{{\sst
[N]}\times{\sst [N]}}\ .
\label{udef}\end{equation}
In turn, the classical ADHM gauge field $A_n$ is constructed from $U$ as
follows:
\begin{equation}A_n{}_{\sst [N] \times [N]}  =
\bar U_{\sst [N] \times \underline{\sst [N+2k]}}\partial_{n}
U_{\underline{\sst [N+2k]} \times [N]} \ .
\label{vdef}\end{equation}
Note first that
for the special case $k=0,$  the 
gauge configuration $A_n$  defined by \eqref{vdef}
is a ``pure gauge'' so that it automatically
solves the selfduality equation \eqref{asdeq}
in the vacuum sector. The ADHM ansatz is that Eq.~\eqref{vdef} continues
to give a solution to Eq.~\eqref{asdeq}, even for nonzero $k$. As we
shall see, this requires two additional conditions. The first one is the
so-called {\it factorization} condition:
\begin{equation}
\bar\Delta_{\sst [2] \times[k]
\times \underline{\sst [N+2k]}}
\Delta_{\sst \underline{[N+2k]} \times [k] \times[2]}=
1_{\sst [2] \times [2]} f^{-1}_{\sst [k]\times [k]} \ ,
\label{dbd}
\end{equation}
where $f$ is an arbitrary $x$-dependent $k\times k$ dimensional
Hermitian matrix.  The second condition is the so-called
{\it completeness} relation:
\begin{equation}
1_{\sst [N+2k] \times[N+2k]}=
\Delta_{\sst [N+2k] \times \underline{\sst [k]} \times \underline{\sst [2]}}
f_{\underline{\sst [k]} \times \underline{\sst [k]} }
\bar\Delta_{\sst \underline{[2]} \times \underline{[k]} \times [N+2k]}+
U_{\sst [N+2k] \times \underline{\sst [N]}}
\bar U_{\sst \underline{[N]} \times [N+2k]} \ .
\label{cmpl}
\end{equation}
Note that both terms on the right hand side of \eqref{cmpl} are Hermitian
projection operators
\EQ{P_{[N+2k]\times[N+2k]}\
\equiv
\Delta_{\sst [N+2k] \times \underline{\sst [k]} \times \underline{\sst [2]}}
f_{\underline{\sst [k]} \times \underline{\sst [k]} }
\bar\Delta_{\sst \underline{[2]} \times \underline{[k]} \times [N+2k]}
\ , \quad
{\cal P}_{[N+2k]\times[N+2k]}\ \equiv
U_{\sst [N+2k] \times \underline{\sst [N]}}
\bar U_{\sst \underline{[N]} \times [N+2k]}
\ . \label{ppdef}
}
Both conditions, \eqref{dbd} and \eqref{cmpl}, will be investigated below
in \S\ref{sec:S22}.
With the above relations together with integrations by parts, the expression
for the field strength $F_{mn}$ for the SD instanton
may  be elaborated as follows:
\begin{equation}\begin{split}
F_{mn} &\equiv \partial_{[m}A_{n]} +  A_{\,[m} A_{n]} =
\partial_{[m}(\bar U\partial_{n]}U)
+(\bar U\partial_{\,[m}U)(\bar U\partial_{n]}U)=
\partial_{\,[m}\bar U(1-U\bar U)\partial_{n]}U
\\&= \partial_{\,[m}\bar U\Delta f \bar\Delta\partial_{n]}U=
\bar U\partial_{[m}\Delta f \partial_{n]}\bar\Delta U =
\bar Ub \sigma_{[m}\bar\sigma_{n]}f \bar{b} U  =
4\bar U b \sigma_{mn}f\bar b U\ .
\label{sdu}\end{split}\end{equation}
Selfduality of the field strength then follows automatically from
\eqref{sigsd}.

For the ASD instanton field strength we get:
\begin{equation}
F_{mn} =
\bar U\partial_{[m}\Delta f \partial_{n]}\bar\Delta U =
\bar Ub \bar\sigma_{[m}\sigma_{n]}f \bar{b} U  =
4\bar U b \bar\sigma_{mn}f\bar b U\ ,
\label{asdu}\end{equation}
which is anti-selfdual due to \eqref{sigasd}.

\subsection{Constraints and projectors}\label{sec:S22}

We have seen that the ADHM construction for $U(N)$ makes essential use
of matrices of various sizes:
$(N+2k)\times N$ matrices
$U$,  $(N+2k) \times 2k$ matrices
$\Delta$, $a$ and $b$, $k\times k$ matrices $f$, and
$2\times2$ matrices
 $\sigma_{n \alpha \aD}$,
 $\sigmabar^{\aD \alpha}_n,$ $x_{\alpha\aD},$ {\it etc\/}.
Accordingly, we introduce a variety
of index assignments:
\begin{align}\hbox{Instanton number indices\ }[k]:\qquad&1\le i,j,l,\ldots\le
k&\notag \\
\hbox{Color indices\ }[N]:\qquad&1\le u,v,\ldots\le N&\notag \\
\hbox{ADHM indices\ }[N+2k]:\qquad&1\le \lambda,\mu,\ldots\le N+2k&\notag \\
\hbox{Quaternionic (Weyl) indices\ }[2]:\qquad&\alpha,\beta,\aD,
\bD,\ldots=1,2&\notag \\
\hbox{Lorentz indices\ }[4]:\qquad&m,n,\ldots=0,1,2,3&\notag
\end{align}
No extra notation is required for the $2k$ dimensional column index attached
to $\Delta,$ $a$ and $b$, since it can be factored as $[2k]=[k]\times[2]
=j\bD,$ {\it etc.\/},
as in Eq.~\eqref{dlt}. With these index conventions, Eq.~\eqref{dlt}
for the SD instanton reads
\begin{equation}
{\rm SD\, instanton:} \quad \Delta_{\lambda  i \aD}(x)
= a_{\lambda  i \aD}+
b_{\lambda  i}^{\alpha}x_{\alpha\aD}\ ,\qquad
\bar\Delta^{\aD\lambda}_{ i  }(x)
 = \abar^{\aD\lambda}_{ i } +
\bar{x}^{\aD \alpha} \, \bar b^\lambda_{\alpha i}\ ,
\label{del}\end{equation}
while the factorization condition \eqref{dbd} becomes
\begin{equation}
\bar\Delta^{\bD\lambda}_{ i  }
\Delta^{}_{\lambda  j \aD} =
   \delta^{\bD}{}_{\aD}  {(f^{-1})}_{ij}\ .
\label{fac}\end{equation}
Notice that by definition
$ \bar\Delta^{\aD\lambda}_i\equiv (\Delta_{\lambda i\aD})^*.$

\underline{Factorization condition and ADHM constraints}

We can make the canonical form \eqref{canform}
a little more explicit with a
convenient representation of the index set $[N+2k]$.
We decompose each ADHM index $\lambda\in[N+2k]$  into\footnote{The Weyl
index $\alpha$ in this decomposition is raised and lowered with the
$\epsilon$ tensor as always, whereas for the $[N]$ and $[k]$ indices
$u$ and $i$ there is no distinction between upper and lower indices.}
\begin{equation}
\lambda = u + i\alpha\ ,\quad1\le u\le N\ ,\quad1\le i\le k\ ,
\quad\alpha=1,2\ .
\label{rplam}\end{equation}
In other words, the top $N\times2k$ submatrices in Eq.~\eqref{canform}
have rows
indexed by $u\in[N],$ whereas the bottom $2k\times2k$ submatrices have
rows indexed by the pair $i\alpha\in[k]\times[2].$ Equation \eqref{canform}
then becomes
\begin{subequations}
\begin{align}
a_{\lambda  j \aD}&=
a_{(u+i\alpha) j \aD}=
\begin{pmatrix}  w_{u j \aD}\\
(a'_{\alpha\aD})^{ }_{ij}\end{pmatrix}_{\phantom{q}}
\ ,\label{aaa}\\
\bar a^{\aD\lambda}_{j}&= \bar a_{j}^{\aD
(u+i\alpha)}=
\big(\bar w^\aD_{j u}\ \ (\bar a^{\prime\aD\alpha})^{}_{ji}
\big)^{\phantom{T} }_{\phantom{q}}\ ,\label{aab}
\\
b_{\lambda j}^\beta&=\,b_{(u+i\alpha) j}^\beta =
\begin{pmatrix}  0 \\ \delta_{\alpha}^{\ \beta}
\delta_{ij}^{}\end{pmatrix}^{\phantom{T}}_{\phantom{q}}
\ , \label{aac}\\
\bar b_{\beta j}^\lambda&= \bar b^{u+i\alpha}_{\beta j}=
\big(0 \ \ \delta^{\ \alpha}_{\beta}\, \delta_{ji}^{}
\big)^{\phantom{T}}\ .
\label{aad}\end{align}
\end{subequations}
Combining Eqs.~\eqref{del}-\eqref{aad}, and noting that $f_{ij}(x)$ is
arbitrary,
one extracts the $x$-independent conditions on the matrix $a$:
\begin{subequations}
\begin{align}
{\rm SD\, instanton:} \quad
&\tau^{c\,\aD}{}_{\bD}(\bar a^\bD a_\aD)_{ij}
-\delta_{ij} \bar\eta^c_{mn}\th^{mn}= 0
\label{fconea}\\
&(a^{\prime}_n)^\dagger  = a^{\prime}_n\ .
\label{fconeb}\end{align}
\end{subequations}
In Eq.~\eqref{fconea} there are three separate equations since
we have contracted $\bar a^\bD a_\aD$
with any of the
three Pauli matrices, while in Eq.~\eqref{fconeb} we have decomposed
 $(a'_{\alpha\aD})_{ij}$ and $(\bar a^{\prime\aD\alpha})_{ij}$
in our usual quaternionic basis of spin matrices:
\begin{equation}(a'_{\alpha \aD})^{}_{ij}=
(a'_n)^{}_{ij}\sigma_{n \alpha\aD}\ , \quad
(\bar a^{\prime\aD \alpha})^{}_{ij} =
(a'_n)^{}_{ij}\bar\sigma_n^{\aD \alpha}\ .
\label{dec}\end{equation}
The three conditions \eqref{fconea}
are the modified ADHM constraints for the SD instanton.
When $\th^{mn}=0$ Eqs.~\eqref{fconea} give the standard
commutative ADHM constraints \cite{CFGT,CWS}.
When noncommutativity is present, the
SD instanton constraints
are modified by the ASD component  of $\th$.
Thus, the ADHM constraints for the SD instanton in the SD-$\th$ background
on  $\bf{R}_{\rm NC}^4$ are unmodified,
\EQ{
\tau^{c\,\aD}{}_{\bD}(\bar a^\bD a_\aD)_{ij}= 0 \ .
\label{sdiasdt}}
At the same time, the constraints for the SD instanton in noncommutative space
$\bf{R}_{\rm NC}^2\times \bf{R}^2$ are modified,
\EQ{
\tau^{c\,\aD}{}_{\bD}(\bar a^\bD a_\aD)_{ij}
-\delta_{ij} \delta^{c3} \zeta= 0 \ .
\label{sdincs}}

The ASD instanton constraints follow from solving
the same factorization condition \eqref{dbd} with the 
matrix $\Delta$
given by \eqref{dltasd}. In this case the ADHM constraints 
are modified by
the SD component  of $\th$, 
\AL{
{\rm ASD\, instanton:} \quad
&\tau^{c\,\aD}{}_{\bD}(\bar a^\bD a_\aD)_{ij}
-\delta_{ij} \eta^c_{mn}\th^{mn}= 0
\label{asdca}\\
&(a^{\prime}_n)^\dagger  = a^{\prime}_n\ .
\label{asdcb}}
From \eqref{asdca} it follows that
the constraints for the ASD instanton in the SD-$\th$ background
on  $\bf{R}_{\rm NC}^4,$ and for the ASD instanton in noncommutative space
$\bf{R}_{\rm NC}^2\times \bf{R}^2,$ are modified in the same way:
\EQ{
\tau^{c\,\aD}{}_{\bD}(\bar a^\bD a_\aD)_{ij}
-\delta_{ij} \delta^{c3} \zeta= 0 \ .
\label{asdic}}

The ADHM constraints define a set of coupled quadratic conditions
on the matrix elements of $a$ which have to be solved in order to determine
each SD $k$-instanton solution explicitly.
The elements of the matrix $a$ comprise the collective coordinates
for the $k$-instanton gauge configuration. Clearly the number
of independent such elements grows as $k^2$, even after accounting
for the constraints. In contrast, the
number of physical collective coordinates should equal $4kN$ which scales
linearly with $k$. It follows that
$a$ constitutes a highly redundant set of parameters. 
Much of this redundancy
can be eliminated by noting that the ADHM construction
with $b$ in the canonical form \eqref{canform}
is unaffected by $x$-independent transformations of the form:
\begin{equation}
\Delta_{\sst [N+2k]\times [2k]}\to
\begin{pmatrix} 1_{\sst [N]\times [N]} & 0_{\sst [2k]\times [N]} \\
0_{\sst [N]\times [2k]}  &
{\cal U}^\dagger_{\sst [2k]\times [2k]}\end{pmatrix}
\Delta_{\sst [N+2k]\times [2k]}{\cal U}_{\sst [2k]\times [2k]}
\label{res}\end{equation}
where ${\cal U}_{\sst [2k]\times [2k]}
 =  \grp_{ij} \delta_{\ \aD}^\bD$
and $\grp_{ij} \in \U(k)$. In terms of $w$ and $a'$, this $U(k)$ symmetry
transformation acts as
\begin{equation}w_{ui\aD} \to w_{uj\aD}\grp_{ji}
\ , \qquad
(a'_{\alpha\aD})_{ij} \to \grp^\dagger_{ik}
(a'_{\alpha \aD})_{kl} \grp_{lj}\ .
\label{restw}\end{equation}

From now on, we will take the basic ADHM variables to be the
complex quantities $w_{ui\aD}$, where $\bar
w^\aD_{iu}\equiv (w_{ui\aD})^*$, and the four $k\times k$ Hermitian
matrices $a'_n$. Now we can count the independent collective
coordinate degrees of freedom of the ADHM $k$-instanton. The matrix
$w_{ui\aD}$ contributes $4kN$ real degrees of freedom and
Hermitian matrices $a'_n$ give $4k^2.$ The ADHM conditions
\eqref{asdic} impose $3k^2$ real constraints, and modding out by
the $U(k)$ symmetry group removes further $k^2$ degrees of
freedom. In total we have precisely $4Nk$ real degrees of freedom
left, which is precisely the expected 
number of independent $k$-instanton
collective coordinates. Of these it is easy to identify four
coordinates which correspond to instanton translations $X_n$: \SP{
{\rm SD\, instanton:} \qquad a_{\sst [N+2k] \times [k] \times [2]}
&= b_{\sst [N+2k] \times [k] \times \underline{\sst [2]}} \
X_{\sst \underline{[2]} \times [2]} \ , \\
{\rm ASD\, instanton:} \qquad
a_{\sst [N+2k] \times [k] \times [2]} &=
b_{\sst [N+2k] \times [k] \times \underline{\sst [2]}} \
\bar X_{\sst \underline{[2]} \times [2]} \ .
\label{trasd}}
as is obvious from \eqref{dlt} and \eqref{dltasd}.

\underline{Completeness relation}

We can now study the completeness relation \eqref{cmpl}. This relation is
automatic in the standard commutative case, but we will point out that there are
subtleties in the noncommutative case, where $x$ itself is an operator.
In the noncommutative case the Hermitian projector
 $P=\Delta f\bar\Delta$ defined in
\eqref{ppdef} is an 
$[N+2k] \times[N+2k]$
matrix of operators on a Fock space $\cH$
\EQ{P\, : \, \cH^{N+2k} \rightarrow P\cH^{N+2k}\subset \cH^{N+2k} \ .
}
We start by considering the eigenvalue problem for $P$.
Since $P$ is a Hermitian
projection operator, $P^\dagger=P$ and $PP=P$, all its eigenvalues
are either equal to zero or equal to one. Let $|\Psi^p\rangle$
and $|\Phi^r\rangle$ denote the normalized zero-mode
and non-zero-mode eigenstates
of P:
\AL{ P |\Psi^p\rangle=0 \ , \quad &|\Psi^p\rangle \in \cH^{N+2k} \ , \quad
\langle \Psi^p|\Psi^q\rangle= \delta^{pq} \ , \label{zmsp}\\
P |\Phi^r\rangle =|\Phi^r\rangle\ , \quad &|\Phi^r\rangle \in
\cH^{N+2k} \ , 
\quad
\langle \Phi^r|\Phi^s\rangle= \delta^{rs} \ . \label{nzmsp}
}
Now, since the set of all eigenstates of a Hermitian operator is complete
we can write
\SP{1_{\sst [N+2k] \times[N+2k]}&=\sum_p |\Psi^p\rangle\langle \Psi^p| +
\sum_r |\Phi^r\rangle\langle \Phi^r| \\
&=\sum_p |\Psi^p\rangle\langle \Psi^p| +
\Delta_{\sst [N+2k] \times \underline{\sst [k]} \times \underline{\sst [2]}}
f_{\underline{\sst [k]} \times \underline{\sst [k]} }
\bar\Delta_{\sst \underline{[2]} \times \underline{[k]} \times [N+2k]}
\ . \label{cmpl2}
}
The second line in \eqref{cmpl2} follows from the fact that all non-zero
eigenvalues of $P=\Delta f\bar\Delta$ are equal to one.

The ADHM relation \eqref{cmpl} will follow from \eqref{cmpl2} if
and only if all the zero-mode eigenstates 
of $P$ 
can be written as: \EQ{
\Bigl\{|\Psi^p\rangle\Bigr\}=\left\{ U_{\sst [N+2k] \times
\underline{\sst [N]}} |s_{\underline{\sst [N]}}\rangle\right\} \ ,
\label{reqt} } where $|s_{u}\rangle$ are arbitrary normalized
states in  $\cH$. If the requirement \eqref{reqt} holds, then
$\sum_p |\Psi^p\rangle\langle \Psi^p|=U\bar U$ and the
completeness relation \eqref{cmpl} follows. If \eqref{reqt} does
not hold, we cannot use \eqref{cmpl} and the ADHM construction of
(A)SD field strengths necessarily breaks down.

Note that the condition \eqref{reqt} is not automatic. While it is
always true that a state $U|s\rangle\neq 0$ is necessarily a normalized
zero-mode eigenstate of $P=\Delta f\bar\Delta$ (this follows from
\eqref{uan}-\eqref{udef}), it is not generally correct to assume
that each zero mode can be represented in this way.
In the following sections we will analyse \eqref{reqt} and \eqref{cmpl}
for various explicit noncommutative instanton solutions. 

\subsection{Topological charge and Corrigan's identity}\label{sec:S23}

The topological charge of the SD ADHM $k$-instanton is given by
\EQ{
Q=-\frac1{16\pi^2}\int d^4x\,{\rm Tr}_NF_{mn}{}{F}_{mn}
\ ,
\label{topc2}
}
where the SD field strength is given by
\EQ{
F_{mn} =
\bar Ub \sigma_{[m}\bar\sigma_{n]}f \bar{b} U
\ .
\label{sdu2}}
The integral in \eqref{topc2} can be evaluated in general thanks 
to a remarkable
Corrigan's identity:
\EQ{{\rm Tr}_NF_{mn}{}{F}_{mn}=
\hf \partial_n\partial^n {\rm Tr}b \sigma_{m}({\cal P}+1)\bar\sigma_{m}bf
\ , \label{coridf}
}
where ${\rm Tr}$ means a trace over both $U(N)$ and ADHM indices
and ${\cal P}=1-\Delta f\bar\Delta$ was defined in \eqref{ppdef}. 
The relation \eqref{coridf} was first derived in the commutative case
in \cite{Corrigan:1979di}. 
A brute-force proof
of \eqref{coridf} in the commutative $SU(2)$ case 
which appeared in
Appendix C2 of \cite{MO1} can be directly applied to the noncommutative
$U(N)$ construction.

Thanks to \eqref{coridf} $Q$ is the integral of a total derivative,
hence the straightforward way to evaluate it is to map the operator-valued
expressions to the operator symbols and saturate the integral on the boundary.
As always
instanton configurations which are 
relevant to semiclassical functional integral
applications are either regular or are gauge equivalent to the regular
configurations. Thus, since \eqref{coridf} is gauge invariant due to
${\rm Tr},$ we can assume that the expression
$\hf \partial^n {\rm Tr}b \sigma_{m}({\cal P}+1)\bar\sigma_{m}bf$
contains no singularities at finite values of $x$. Hence, $Q$
receives contributions solely from the  boundary $S^3$ at infinity,
\EQ{
Q=\frac1{16\pi^2}\frac1{2} 2\pi^2\cdot2 \ {\rm Tr}\ b \sigma_{m}
(\Pinfty+1)\bar\sigma_{m} b\ = k
\ . \label{qcalc}}
In deriving \eqref{qcalc} we used the following asymptotics:
\begin{equation}\label{Pinftydef}
\Delta \rightarrow bx \ , \quad 
f_{ij} \rightarrow {1\over x^2}\delta_{ij} \ , \quad
\P \rightarrow  1-b\bbar \equiv \Pinfty\ , \quad\quad
\mbox{ as $|x|\rightarrow\infty$}.
\end{equation}
Thus, we conclude that the topological charge of the noncommutative
SD ADHM $k$-instanton
is always equal to $k$. An almost identical calculation for the ASD
$k$-instanton gives $Q=-k.$
It is remarkable that the fact that the topological charge is an integer
and is equal to $\pm k$
is basically an algebraic statement encoded
in the structure of the ADHM matrices  even in the noncommutative case.
We would like to stress that this general result is independent of the rank
of $\theta$ and
applies equally well
to the case of space-space noncommutativity. 
In the following sections we will evaluate topological charges of
some simple instanton solutions without making use of this powerful argument.

\section{U(1) single-instanton solution}\label{sec:S3}

In this section we analyse in detail an
explicit construction of the ASD single
instanton solution in the noncommutative $U(1)$ gauge theory. Singular
$U(1)$ instantons on $\bf{R}_{\rm NC}^4$
were first discussed in \cite{NS}, and the regular solutions
were constructed in \cite{Furuuchi1,Furuuchi2}. Here we will treat
the two noncommutative backgrounds: (1) SD-$\th$ on $\bf{R}_{\rm NC}^4,$
and (2)
$\bf{R}_{\rm NC}^2\times \bf{R}^2$
in parallel.
In case (1) regular ADHM solutions will be constructed with the use of a shift
operator $u^\dagger$ required to project out certain states from 
the Hilbert space, while in case (2) we will see that
one should not project out any states.

The ADHM matrix $\Delta$ for the ASD instanton \eqref{dltasd}
which satisfies the modified ADHM constraints
\eqref{asdcb}, \eqref{asdic} is given by:
\EQ{\Delta=\begin{pmatrix}\sqrt{\zeta}&\, & 0\\ \zb_2-\Zb_2 &\, & -(z_1-Z_1) \\
\zb_1-\Zb_1 &\, & z_2-Z_2 \end{pmatrix} \ , \qquad
\bar\Delta=\begin{pmatrix}\sqrt{\zeta}&\, & z_2-Z_2&\, &z_1-Z_1 \\
0 &\, & -(\zb_1-\Zb_1)&\, & (\zb_2-\Zb_2) \end{pmatrix} \ .
\label{del1u1}} The expressions above are written in the complex
coordinates $z_1,z_2,\zb_1,\zb_2$ basis \eqref{rae}. The
translational collective coordinates of the instanton,
$Z_1,Z_2,\Zb_1,\Zb_2,$ are c-numbers. Equation \eqref{del1u1}
gives the general solution to the constraints
\eqref{asdcb},\eqref{asdic}, and the $U(1)$ instanton moduli space
is simply the $\bf{R}^4$ which is spanned by $Z_1,Z_2,\Zb_1,\Zb_2,$ or
equivalently, $X_n$ of \eqref{trasd}. From now on  we will
always set these overall translations of the instanton to zero,
$Z_1=Z_2=\Zb_1=\Zb_2=0.$

The factorization condition \eqref{dbd}
is then automatically satisfied and
\EQ{ f={1\over \zeta +z_1\zb_1 +z_2\zb_2} \ . \label{f1u1}}
The final step 
in the ADHM set-up is the construction of the normalized matrix $U$
such that $\bar\Delta U=0$ and $\bar U U=1$, as required by
\eqref{uan},\eqref{udef}, and the expression for the gauge field will follow
from \eqref{vdef}.
The (unnormalized) solution $\tilde U_0$ is easy to find:
\EQ{\tilde U_0= \begin{pmatrix} z_1\zb_1 +z_2\zb_2\\ -\sqrt{\zeta}\zb_2\\
-\sqrt{\zeta}\zb_1 \end{pmatrix} \ , \quad \bar\Delta \tilde U_0=0 \ .
\label{ut01u1}
}
The problem is that $\tilde U_0$ is not straightforwardly normalizable.

\subsection{ $\bf{R}_{\rm NC}^4$}\label{sec:S41}
Let us start with $\bf{R}_{\rm NC}^4$ space with  SD
$\th$. It is easy to see that $\tilde U_0$ annihilates the
vacuum,\footnote{ For an instanton centered at
$(Z_1,Z_2,\Zb_1,\Zb_2)$ the annihilated state will be the
corresponding coherent state $|\Zb_1,\Zb_2\rangle.$} $\tilde
U_0|0,0\rangle = 0.$ In order to find the normalized expression
for $U$, the vacuum state $|0,0\rangle$ has to be projected out.
An elegant realization of this idea was proposed in
\cite{Furuuchi2}. First define a projector $p = 1-|0,0\rangle
\langle 0,0|.$ Then the normalized matrix $U$ can be determined via
\EQ{
U=\tilde U_0 \beta_p u^\dagger \ , \quad \bar U  U=1 \ ,
\label{udeff} } 
where $\beta_p$ is the normalization factor,
\EQ{
\beta_p=p {1\over\sqrt{\tilde U_0^{\dagger}\tilde U_0}}p=
p{1\over\sqrt{(z_1\zb_1 +z_2\zb_2)(z_1\zb_1 +z_2\zb_2+\zeta)}} p\ ,
\label{betdef}} 
and $u^\dagger$ is a shift operator which projects
out the vacuum: \SP{&u^\dagger \ : \ \cH \rightarrow p\cH \ ,
\quad
u \ : \ p\cH \rightarrow \cH \ , \label{udagdef} \\
&u u^\dagger = 1 \ , \quad u^\dagger u = p\ , \quad p u^\dagger =u^\dagger \ ,
\quad u p=u \ .
}
Due to the factors of $p$ in the definition of $\beta_p$
which projects out the dangerous vacuum state, the right hand side of
\eqref{betdef} is not singular and well-defined.

It is also straightforward to check that the ADHM completeness
relation \eqref{cmpl} is satisfied, $1-U\bar U=\Delta
f\bar\Delta,$ and the field-strength $F_{mn}$ is in fact
anti-selfdual. The topological charge $Q$ can be now calculated
as a trace on the Hilbert space. The result is \cite{NS,Furuuchi2}
\EQ{ Q=-4 \sum_{(n_1,n_2)\neq 0} {1\over
(n_1+n_2)(n_1+n_2+1)^2(n_1+n_2+2)}= -4 \sum_{N=1}^\infty{1\over
N(N+1)(N+2)}= -1 \ , \label{q11fur}} in agreement with the general
argument of the previous section \S\ref{sec:S23} that $|Q|=k.$ An
alternative calculation from \cite{Furuuchi2} evaluates
$Q=-1$ by integrating over the c-number operator symbols.

To summarize: the gauge-field ASD
instanton configuration resulting from \eqref{udeff},
is a local minimum
of the action in the $U(1)$ theory on $\bf{R}_{\rm NC}^4$
with the SD-$\th$. The instanton action is $S_{\rm inst}=8\pi^2/g^2,$
and the topological charge is $Q=-1.$ The instanton configuration
is perfectly regular and can be expanded around in the functional
integral, leading to quantum instanton contributions
in the $U(1)$ theory in the SD-$\th$ background.
At the same time the self-dual ADHM
$U(1)$ instanton in a  SD-$\th$ background does not
exist as the corresponding unmodified ADHM constraints have no non-trivial
solution for $N=1.$

\subsection{$\bf{R}_{\rm NC}^2\times\bf{R}^2$}\label{sec:S42}
 We now consider $\bf{R}_{\rm NC}^2\times\bf{R}^2$ space.
The matrix $\tilde U_0$ in \eqref{ut01u1}
annihilates the vacuum state $|0\rangle$ of $\bf{R}_{\rm NC}^2$
{\it at the point}
$z_2=0=\zb_2.$ This is the crucial difference from the unconditional
annihilation of a state in the previous example. 

Let us first try
to follow the same route as in 
\S\ref{sec:S41} and normalize $U$ by projecting
out the offending state. We will see momentarily that this approach
will fail since 
the completeness relation will be violated, leading to a
gauge configuration which is not anti-self-dual. 
However, it will turn out that on $\bf{R}_{\rm NC}^2\times\bf{R}^2$
regular anti-self-dual instantons can be constructed
without projecting out any states.

To see this
we first introduce
a projector $p = 1-|0\rangle \langle 0|,$ and define the normalized
matrix $U$ via \eqref{udeff}-\eqref{udagdef}.
A straightforward calculation shows that the ADHM completeness relation
\eqref{cmpl} is not satisfied, and, hence, the field-strength
$F_{mn}$ is not anti-selfdual.
In fact, let us check the relation,
$1-U\bar U=\Delta f\bar\Delta,$ for the $11$-matrix element:
\SP{(1-U\bar U)_{11}&=1-\sum_{n\neq 0} { z_1\zb_1 +z_2\zb_2 \over
z_1\zb_1 +z_2\zb_2+\zeta} |n\rangle\langle n| \ , \\
(\Delta f\bar\Delta)_{11}&={\zeta\over z_1\zb_1 +z_2\zb_2+\zeta} \ , \\
(1-U\bar U)_{11}-(\Delta f\bar\Delta)_{11}&=
{z_2\zb_2\over z_2\zb_2+\zeta}|0\rangle\langle 0| \ . }
The last line is non-zero everywhere 
except at the origin $z_2 = \zb_2 =0$. 
This, of course, invalidates the ADHM construction.
Since this point is important,  it is worthwhile to see more explicitly 
how the sum of the two orthogonal projectors $P$ and $\cP$ fails to 
span the whole Hilbert space. 
Consider the normalized state
\EQ{
\ket{\psi} = \begin{pmatrix} \ket{0}\cr 0 \cr 0
\end{pmatrix} .
}
It is easy to check that 
\EQ{ \label{projs}
\cP \ket{\psi} =0,
\quad
P \ket{\psi} \neq  \ket{\psi}.
}
In fact
\EQ{
P \ket{\psi} =  \begin{pmatrix} 
{\z (\z + z_2\zb_2)^{-1}} \ket{0} 
\cr {\sqrt{\z} \zb_2 (\z + z_2\zb_2)^{-1}} \ket{0} \cr 0
\end{pmatrix} 
\neq \ket{\psi} \quad \mbox{unless $z_2 =0=\zb_2$}.
} 
Therefore we see that the  orthogonal projectors $P$ and $\cP$ do not span
the whole Hilbert space.

Now we ask what happens if we do not subtract the vacuum state.
In this case the ADHM matrix $\tilde{U}_0$ is
normalizable for all values of
$z_2$ and $\zb_2$ except at the origin $z_2=0=\zb_2.$ The resulting gauge
field configuration is singular at $z_2=0=\zb_2.$
This singularity is of the form
$\phi_{\rm sing}(x_3,x_4) \cdot |0\rangle\langle 0|$ and
is much
more severe than the well-known point-like singularity of the
commutative instanton in the singular gauge. The c-number operator symbol
gauge field will contain a term
$\phi_{\rm sing}(x_3,x_4)e^{-(x_1^2+x_2^2)/ \zeta},$ which is singular
at $x_3=0=x_4$ on the whole $(x_1,x_2)$-plane,
as follows from \eqref{fsos}. 
The idea that this singularity is a gauge artifact 
was recently put forward in \cite{K2} based on the observation
that ${\rm Tr} F^n$ is non-singular. 

We will now prove explictly that this singularity
can be removed by a singular gauge transformation $g$
and a regular instanton can indeed be constructed. 
The normalized ADHM matrix $U$ which gives a regular U(1) instanton
reads
\EQ{ U = \tilde U_0 \beta g^\dagger\ , \qquad \bar{U} U=1 \ ,
\label{newu1u1}
}
where $\tilde U_0$ is given by \eqref{ut01u1}, 
$\beta$ is the normalization factor
\EQ{
\beta= 
{1\over\sqrt{(z_1\zb_1 +z_2\zb_2)(z_1\zb_1 +z_2\zb_2+\zeta)}} \ ,
\label{newbetdef}} 
and  
\EQ{g^\dagger = |0\rangle \langle 0| {z_2 \over |z_2 | } + 
\sum_{n >0} | n \rangle \langle n |
\ .  \label{gdes}
}  
Note an important difference with \eqref{udeff}:
no shift operator has been introduced in \eqref{newu1u1},
$g^\dagger$ is a true gauge transformation,
\EQ{g^\dagger g =1 = g g^\dagger \ , }
and no states have been projected out
from the Hilbert space. Consequently, the definition of $\beta$
\eqref{newbetdef} does not contain any projector.

The corresponding ASD field strength is now easily computed using 
\eqref{asdu} and is given by
\begin{eqnarray}
\label{F1234}
F &=& g \beta \zeta
\left[(z_2 f \zb_2-z_1 f \zb_1)
(dz_1 d\bar{z}_1 - dz_2 d\bar{z}_2)
+2z_1 f \zb_2 d\zb_1 dz_2 + 2 z_2 f \zb_1 d\zb_2 dz_1\right] 
\beta g^\dagger \\ \nonumber
&\equiv& g F' g^\dagger
\ , 
\end{eqnarray}
with $f$ defined in \eqref{f1u1}.
Note that the operator $F'$ is singular\footnote{
$F'$ was first written down in \cite{K2}.}, in fact
$<0| F'_{2 \bar 1}$ and  $F'_{1 \bar 2}|0>$ 
are not well defined at $z_2=0=\zb_2$. 
The singular parts
are given by
\EQ{
F^{\prime \, \rm sing}_{\bar{2}1} =
{\sqrt{2}\over \zeta} {z_2 \over |z_2|}\, |0\rangle \langle 1| \ ,
\quad 
F^{\prime \, \rm sing}_{\bar{1}2} =
{\sqrt{2}\over \zeta} {\bar z_2 \over |z_2|}\, |1\rangle \langle 0| \ .
}
Correspondingly, the leading singularity in the gauge field one-form
is of the type
\EQ{A^{\prime \, \rm sing} = {z_2 \over |z_2|}
d\left({\zb_2 \over |z_2|}\right)
\,|0\rangle \langle 0| \ .
}
The role of $g$ is now clear: it is a singular gauge transformation which
removes singularities from $F^{\prime \, \rm sing}$
and $A^{\prime \, \rm sing},$
\EQ{
gF^{\prime \, \rm sing}_{\bar{2}1}g^\dagger =
{\sqrt{2}\over \zeta}  |0\rangle \langle 1| \ ,
\quad 
gF^{\prime \, \rm sing}_{\bar{1}2} g^\dagger=
{\sqrt{2}\over \zeta}  |1\rangle \langle 0| \ .
} 
At the same time no new singularities are introduced. 
Hence, we have constructed a regular
instanton solution with a space-space noncommutativity.

The topological charge of this instanton is guaranteed
to be equal to minus one as a straightforward
consequence of the Corrigan's identity \eqref{qcalc}.

\section{U(1) two-instanton solution}\label{sec:S4}

\subsection{$\bf{R}_{\rm NC}^4$}\label{sec:Snew41}

In this section we study the ADHM $2$-instanton solution in the
$U(1)$ gauge theory on $\bf{R}_{\rm NC}^4.$
The general $2$-instanton solution was first studied in \cite{LTY}.
The ADHM matrix $\bar\Delta$ for the ASD instanton \eqref{dltasd}
which satisfies the modified
ADHM constraints \eqref{asdcb},\eqref{asdic}
is given by:
\EQ{
\bar\Delta=\begin{pmatrix}\sqrt{\zeta}\sqrt{1-b} & z_2-\delta_2 &
-\delta_2\sqrt{2b\over a} & z_1-\delta_1 & -\delta_1\sqrt{2b\over a}\\
\sqrt{\zeta}\sqrt{1+b} & 0 & z_2+\delta_2 & 0 & z_1+\delta_1 \\
0& -(\zb_1-\delta_1^*) & 0 & \zb_2-\delta_2^*& 0 \\
0& \delta_1^*\sqrt{2b\over a}& -(\zb_1+\delta_1^*) &
-\delta_2^*\sqrt{2b\over a} &\zb_2+\delta_2^* \end{pmatrix} \ ,
\label{del2u1}}
where $\delta$'s are arbitrary c-numbers and
\EQ{a={2\over\zeta}(|\delta_1|^2+|\delta_2|^2) \ , \quad
b=\sqrt{1+a^2}-a \ .
\label{abdefs}}
Equation \eqref{del2u1} gives the general solution of the constraints
\eqref{asdcb},\eqref{asdic}, with the center of mass collective
coordinates set to zero, $Z_1=Z_2=\Zb_1=\Zb_2=0.$
The unconstrained collective coordinates $\delta_i$ and $\delta_i^*$
give the center of the first instanton; the second
instanton is centered at $(-\delta_i,-\delta_i^*).$ The $2$-instanton
moduli space is $8$-dimensional as required, and is spanned by
four $Z$'s and four $\delta$'s.
As shown in \cite{LTY}, after separating the center of mass,
the relative moduli space is given by the Eguchi-Hanson manifold
(which is non-singular even at the origin where the two point-like
$U(1)$ instantons coincide).

In the dilute instanton gas limit $|\delta_i|\to \infty,$ the expression
on the right hand side of \eqref{del2u1} clearly splits into two
single-instanton expressions:
\EQ{
\bar\Delta_{\infty}=
\begin{pmatrix}\sqrt{\zeta} & z_2-\delta_2 &
0 & z_1-\delta_1 & 0\\
0 & 0 & 0 & 0 & 0 \\
0& -(\zb_1-\delta_1^*) & 0 & \zb_2-\delta_2^*& 0 \\
0& 0& 0& 0\end{pmatrix} +
\begin{pmatrix}0&0&0&0\\
\sqrt{\zeta} & 0 & z_2+\delta_2 & 0 & z_1+\delta_1 \\
0& 0&0&0\\
0& 0& -(\zb_1+\delta_1^*) &
0 &\zb_2+\delta_2^* \end{pmatrix} \ .
\label{deldg}}
In the opposite limit of coincident instantons, $\delta_i\to 0,$ and
$a\to 0,$ $b\to 1$ in such a way that
\EQ{\delta_1\sqrt{2b\over a} \to -\sqrt{\zeta}\lambda_1 \ , \quad
\delta_2\sqrt{2b\over a} \to -\sqrt{\zeta}\lambda_2 \ , \quad
|\lambda_1|^2+|\lambda_2|^2=1 \ .
\label{lamdeff}
}
Here the $\lambda_1$ and $\lambda_2$ are the collective
coordinate which describe the three angles of the direction in which
the two instantons have approached each other.
This coincident $2$-instanton solution was studied in \cite{Furuuchi1}.

We expect that the general $2$-instanton ASD configuration is a regular
solution to the ASD equation which gives a local minimum of the action
$S=16\pi^2/g^2$ and has topological charge $Q=-2$. Then it can contribute
to the functional integral and is semiclassically relevant.
The topological charge of the coincident solution was studied in \cite{KLY}
where it was concluded that $Q$ is not integer and is in general
moduli-dependent. If true, this would make the instanton action
moduli-dependent which would conflict with the statement that instantons
are local minima of the action. We have redone the calculation of \cite{KLY}
and found $Q=-2$.

To simplify things a little we fix the angle of approach as in
\cite{KLY}, $\lambda_1=1,$ $\lambda_2=0.$ The corresponding
$\Delta$ and $\bar\Delta$ matrices are:
\EQ{\Delta=\begin{pmatrix}0&\sqrt{2\zeta}& 0 & 0\\
\zb_2& 0 & -z_1 & -\sqrt{\zeta}\\
0& \zb_2& 0 &-z_1\\
\zb_1 & 0 & z_2& 0\\
\sqrt{\zeta}&\zb_1 & 0 & z_2\end{pmatrix} \ , \quad
\bar\Delta=\begin{pmatrix}0 & z_2 &
0 & z_1 & \sqrt{\zeta}\\
\sqrt{2\zeta}& 0 & z_2& 0 & z_1 \\
0& -\zb_1 & 0 & \zb_2& 0 \\
0& -\sqrt{\zeta}& -\zb_1&
0 &\zb_2\end{pmatrix} \ .
\label{dell21}}
The unnormalized matrix $\tilde U_0,$ which solves $\bar\Delta \tilde U_0=0,$
is given by
\EQ{\tilde U_0= \begin{pmatrix}
{1\over \sqrt{2\zeta}}((z_1\zb_1 +z_2\zb_2)(z_1\zb_1 +z_2\zb_2-\zeta/2)
+\zeta z_2\zb_2) \\
\sqrt{\zeta}\zb_2\zb_1\\
-\zb_2(z_1\zb_1 +z_2\zb_2+\zeta/2)\\
\sqrt{\zeta}\zb_1\zb_1\\
-\zb_1(z_1\zb_1 +z_2\zb_2-\zeta/2)\end{pmatrix} \ . \label{ut02u1}
} This expression annihilates two states: $|0,0\rangle$ and
$|1,0\rangle,$ which have to be projected out as in \eqref{udeff},\eqref{udagdef}, 
\EQ{
U=\tilde U_0 \beta_p u^\dagger \ , \quad \bar U  U=1 \ ,
\label{nnudeff} } 
where $\beta_p$ is the normalization factor,
\EQ{
\beta_p
=p{\sqrt{2\z}\over\sqrt{\left[(z_1\zb_1 +z_2\zb_2)(z_1\zb_1 +z_2\zb_2+\z/2)
-\z z_1\zb_1\right]\left[(z_1\zb_1 +z_2\zb_2+\z/2)(z_1\zb_1 +z_2\zb_2+2\z)
-\z z_1\zb_1\right]}} p
\label{nnbetdef}}
and the projector $p$ is
given by $p = 1-|0,0\rangle \langle 0,0|-|1,0\rangle \langle
1,0|.$

The factorization condition \eqref{dbd} is automatically satisfied and
\EQ{ f^{-1}=\begin{pmatrix} \zeta+z_1\zb_1+z_2\zb_2 & \sqrt{\zeta}\zb_1\\
\sqrt{\zeta}z_1& 2\zeta+z_1\zb_1+z_2\zb_2 \end{pmatrix}
 \ , \label{f22u1}}
which can be inverted as follows:
\EQ{ f=\begin{pmatrix} {n_1+n_2+5\over (n_1+n_2+2)(n_1+n_2+5)-2(n_1+1)}&
-{1\over(n_1+n_2+2)(n_1+n_2+5)-2(n_1+1)}\sqrt{2}\zb_1\\
 & \\
-{1\over(n_1+n_2+4)(n_1+n_2+1)-2n_1}\sqrt{2}z_1 &
{n_1+n_2+1\over (n_1+n_2+4)(n_1+n_2+1)-2n_1}
\end{pmatrix}
 \ , \label{finu1}}
where we have set $\zeta=2$ and introduced the
SHO occupation numbers $n_1=z_1\zb_1$ and $n_2=z_2\zb_2.$

We can now evaluate the field strength \eqref{asdu} and represent the
topological charge $Q$ as a trace over $p\cH$.
We have obtained an analytic
expression for the topological charge density identical to the expression
in Eq.~(31) of \cite{KLY}. To determine the topological charge $Q,$ we evaluated
the corresponding trace by summing over the SHO occupation numbers
$(n_1,n_2)\neq (0,0) \neq (1,0).$ We performed this double infinite sum
numerically in Maple sampling over 40,000 points $(n_1,n_2).$
Our result is
\EQ{Q\simeq-1.99987\simeq-2 \ ,}
which is different from the numerical
calculation of \cite{KLY}--versions 1-3 which gave $-0.932.$

\subsection{$\bf{R}_{\rm NC}^2 \times \bf{R}^2$}\label{sec:Snew42}

In order to construct a regular $U(1)$ 2-instanton solution on
$\bf{R}_{\rm NC}^2 \times \bf{R}^2$ we look for an  ADHM matrix $U$
of the form \eqref{newu1u1}
\EQ{ U = \tilde U_0 \beta g^\dagger\ , \qquad \bar{U} U=1 \ .
\label{nnu1u1}
} 
where $\tilde U_0$ is now given by (cf. \eqref{ut02u1})
\EQ{\tilde U_0= \begin{pmatrix}
{1\over \sqrt{2\zeta}}[(z_1\zb_1 +z_2\zb_2)(z_1\zb_1 +z_2\zb_2-\zeta)
+\zeta z_2\zb_2] \\
\sqrt{\zeta}\zb_2\zb_1\\
-\zb_2(z_1\zb_1 +z_2\zb_2)\\
\sqrt{\zeta}\zb_1\zb_1\\
-\zb_1(z_1\zb_1 +z_2\zb_2-\zeta)\end{pmatrix} \ . \label{nnut00}
}
$\beta$ is the corresponding normalization factor and 
$g^\dagger$ is a singular gauge transformation to be determined shortly.
It is easy to see that $\tilde U_0$ annihilates the states
$|0\rangle$ and $|1\rangle$ at the point $z_2=0=\zb_2.$
The gauge transformation $g^\dagger$ will now be determined
from the
singular part of $\tilde U_0 \beta$:
\EQ{\tilde U_0 \beta= \begin{pmatrix}
0 \\
0\\
-{\zb_2\over|z_2|} \left( \ket{0}\bra{0} +{1\over\sqrt{2}}
\ket{1}\bra{1}\right)\\
0\\
0\end{pmatrix}
\, + \, \cdots \ , \label{nnutsg}
}
where the dots stand for terms regular in the $z_2 \to 0$ limit.
Therefore, the  singular gauge transformation which removes this singularity
is given by
\EQ{g^\dagger = {z_2 \over |z_2 | } \left( 
|0\rangle \langle 0|+ |1\rangle \langle 1|\right)+
\sum_{n >1} | n \rangle \langle n |
\ ,  \label{nngdes}
}
and leads via \eqref{nnu1u1} to a regular 2-instanton on 
$\bf{R}_{\rm NC}^2 \times \bf{R}^2$.
Again, the instanton number is equal to minus two 
as a consequence of \eqref{qcalc}.

\section{U(N) instantons }\label{sec:S5}

\subsection{Commutative ASD instanton}\label{sec:S51}

Before addressing noncommutative instantons,
we first recall the ADHM construction of the standard ASD
$1$-instanton solution in commutative $U(N).$
The ASD $1$-instanton is determined from the ADHM matrices \eqref{dltasd}
subject to the constraints \eqref{asdcb} and \eqref{asdic} with
$\zeta\equiv 0$.
Eq.~\eqref{asdcb} says that $a'_n$ is real,
\begin{equation}
a'_n \ \equiv \ -X_n \ \in \ {\bf R}^4\ ,
\label{xcon}\end{equation}
after which Eq.~\eqref{asdic} with $\zeta\equiv 0$ collapses to
\begin{equation}\wbar^\dalpha_u \, w_{u \dbeta} \ = \
\rho^2\,\delta^\dalpha_{\ \dbeta}  \ .
\label{wcon}\end{equation}
The quantities $\rho$ and $X_n$ will soon be identified with the instanton
scale size and space-time position, respectively.
It is convenient to put $w$ in the form:
\begin{equation}w_{\sst [N]\times[2]}
\ = \ \rho\grp_{\sst [N] \times [N]} \
\begin{pmatrix}0_{\sst [N-2] \times [2]} \\
  1_{\sst [2]\times [2]}\end{pmatrix} \ , \qquad
\grp \in {U(N) \over U(1) U(N-2)}\ .
\label{wcon2}
\end{equation}
Setting $\grp=1$ initially, we find for $\Delta$ and $f\,$:
\begin{equation}
\Delta_{\sst [N+2]\times [2]} \ = \
\begin{pmatrix} 0_{\sst [N-2] \times [2]} \\
  \rho \cdot 1_{\sst [2]\times [2]} \\
  y_{\sst [2]\times [2]}\end{pmatrix}\ ,\qquad
f  \ = \ {1 \over y^2 + \rho^2} \ ,
\label{dlsi}\end{equation}
with $y=  x-X$.
We now solve \eqref{uan}, \eqref{udef} 
and determine the normalized matrix $U$:
\begin{equation}
U_{\sst [N+2]\times [N]}  \ = \
\begin{pmatrix} 1_{\sst [N-2]\times [N-2]}& 0 \\
0 & \left( {y^2 \over y^2 + \rho^2} \right)^{1/2}
1_{\sst [2]\times [2]} \\
0_{\sst [2]\times [N-2]}  &
-\left( {\rho^2 \over y^2 (y^2+ \rho^2)} \right)^{1/2}
\bar{y}_{\sst [2]\times [2]}  \end{pmatrix}\ .
\label{ueq}\end{equation}
The gauge field then
follows from Eq. \eqref{vdef}:
\begin{equation}
A_n \ = \ \begin{pmatrix} 0 & 0 \\ 0 & A_n^{\sst SU(2)} \end{pmatrix}
\ .
\label{sutin}\end{equation}
Here
$A_n^{\sst SU(2)}$ is the standard singular-gauge $SU(2)$ anti-instanton 
\cite{BPST,tHooft}
with  space-time position $X_n\,$, size $\rho$ and
in a  fixed iso-orientation:
\begin{equation}
 A_n^{\sst SU(2)} (x)\ = \ {i \rho^2 \ \eta^a_{nm}\ (x-X)^m \ \tau^a
\over (x-X)^2 \ ((x-X)^2 + \rho^2)}
\ .
\label{singin}\end{equation}
For a general iso-orientation matrix $\grp$ we obtain instead
\begin{equation}
A_n \ = \ \grp \
\begin{pmatrix} 0 & 0 \\ 0 & A_n^{\sst SU(2)}\end{pmatrix}{\grp}^\dagger
\ , \qquad
\grp \in {U(N) \over U (1) \times U(N-2)}\ .
\label{inss}\end{equation}
We see that the instanton always lives in an $SU(2)$ subgroup of the $SU(N)$
gauge group. An explicit representation of this embedding is formed by
the three composite $SU(2)$ generators
\begin{equation}
\left(t^c\right)_{uv}\  =\ \rho^{-2}\,w^{}_{u\aD}
\left(\tau^c\right)^\aD_{\ \bD}\bar
w_{v}^\bD,\qquad c=1,2,3.
\label{embed}
\end{equation}

\subsection{ASD instanton in the SD background on 
$\bf{R}_{\rm NC}^4$}\label{sec:S52}

The ASD $1$-instanton in the SD-$\th$ background in the noncommutative
$U(N)$ theory is characterized by the ADHM matrices \eqref{dltasd}
subject to the constraints \eqref{asdcb} and \eqref{asdic}.
For the case at hand with $k=1$ these constraints are solved by
choosing the $N\times 2$ matrix $w$ in \eqref{aaa} in the form:
\EQ{w_{\sst[N]\times[2]} = \grp_{\sst[N]\times[N]}
\begin{pmatrix} 0_{\sst[N-2]} & 0_{\sst[N-2]}\\
0&\rho\\ \sqrt{\zeta+\rho^2}&0\end{pmatrix} \ , \quad
\grp \in {U(N)\over U(1) U(N-2)} \ ,
\label{wsol}
}
where $\rho$ denotes the instanton size, and $\grp_{\sst[N]\times[N]}$
specifies the embedding of the $U(2)$ subgroup into the gauge group
$U(N)$. The expression in \eqref{wsol} gives the general solution
of the ADHM constraints.
It follows from \eqref{wsol} that for $N\ge2$
the $U(N)$ noncommutative
instantons are essentially given by the $U(2)$ noncommutative instantons.
The fact that the building blocks for the noncommutative instantons
gauge fields are the $2\times 2$ matrices in the group space, just as
in the ordinary commutative case, is non-trivial. One might have expected
that, since there are noncommutative $U(1)$ instantons, 
two types of building blocks for noncommutative $U(N)$ could exist: 
$U(2)$-instantons and
$U(1)$-instantons. We will see that the $U(1)$-instanton building blocks
appear {\it inside} the $U(2)$ blocks
when the instanton size $\rho$ shrinks to zero.

To keep expressions simple, 
from now on we set  $\grp_{\sst[N]\times[N]}=1.$
In this case the instanton positioned at the origin is determined from:
\EQ{\Delta=\begin{pmatrix}
0_{\sst[N-2]} & 0_{\sst[N-2]}\\0&\rho\\ \sqrt{\zeta+\rho^2}&0 \\
\zb_2 & -z_1 \\
\zb_1 & z_2 \end{pmatrix} \ , \qquad
\bar\Delta=\begin{pmatrix}
0_{\sst[N-2]} & 0 & \sqrt{\zeta+\rho^2}& z_2 & z_1 \\
0_{\sst[N-2]} & \rho & 0& -\zb_1 & \zb_2
 \end{pmatrix} \ .
\label{del1un}} 
Since the instanton configuration is concentrated in a
$U(2)$ factor of the gauge group, we will set $N=2$ from now on. 
The factorization relation follows and
\EQ{f= {1\over z_1\zb_1+z_2\zb_2+\rho^2+\zeta}
\ . \label{fdefnn}}
There are two interesting special cases to notice.
First, when $\zeta \to 0,$ Eqs.~\eqref{del1un}-\eqref{fdefnn}
collapse to the defining equations for the ordinary commutative
BPST instanton (see section 2.3 of \cite{KMS} for a review).
On the other hand, we can consider the limit 
$\rho \to 0,$ with $\zeta$ fixed.
In this case Eqs.~\eqref{del1un}-\eqref{fdefnn}
collapse essentially to the $U(1)$ instanton case \eqref{del1u1}-\eqref{f1u1}.
Thus we conclude that the regular
$U(1)$ instantons arise in the limit of
instanton sizes going to zero.

An ansatz for the unnormalized matrix $\tilde U_0$ was 
given in \cite{Furuuchi1}:
\EQ{\tilde U_0= \begin{pmatrix} 
0 & (z_1\zb_1 +z_2\zb_2 +\zeta)
\sqrt{z_1\zb_1 +z_2\zb_2\over z_1\zb_1 +z_2\zb_2 +\zeta} \\
z_1\zb_1 +z_2\zb_2 & 0 \\
-\sqrt{\zeta+\rho^2}\zb_2 &
\rho z_1\sqrt{z_1\zb_1 +z_2\zb_2\over z_1\zb_1 +z_2\zb_2 +\zeta} \\
-\sqrt{\zeta+\rho^2}\zb_1 &
-\rho z_2\sqrt{z_1\zb_1 +z_2\zb_2\over z_1\zb_1 +z_2\zb_2 +\zeta}
 \end{pmatrix} \ , \quad
\bar\Delta \tilde U_0=0 \ .
\label{ut01un}
}
As in the $U(1)$ case, it is easy to see that $\tilde U_0$ annihilates the
vacuum $\tilde U_0|0,0\rangle = 0,$ which has to be projected out as in
\eqref{udeff},\eqref{udagdef}.
The normalization factor is given by
\EQ{\beta_p= p {1\over\sqrt{\tilde U_0^{\dagger}\tilde U_0}}p =
p \; 1_{\sst [2]\times [2]}{1\over
\sqrt{(z_1\zb_1 +z_2\zb_2)(z_1\zb_1 +z_2\zb_2+\zeta+\rho^2)}} p\ .
\label{betdn}}
However it is easy to check that a $U$ of the form \eq{udeff} 
\EQ{ U= \tilde U_0 \b_p u\dag \ ,
\label{udefff}
}
does not satisfy the completeness condition.
To see this, let us first give the expression $\Delta f
\Delta\dag$. It is 
\EQ{
\Delta f \bar\Delta  = \begin{pmatrix} 
\r^2 f & 0 & -\r f \zb_1 & \r f \zb_2 \\
0 & (\z + \r^2) f & \sqrt{\z + \r^2} f z_2 & \sqrt{\z + \r^2} f z_1 \\
- \r z_1 f & \sqrt{\z + \r^2}  \zb_2 f & \zb_2 f z_2 + z_1 f \zb_1 &
\zb_2 f z_1 - z_1 f \zb_2 \\
\r z_2 f & \sqrt{\z + \r^2}  \zb_1 f & \zb_1 f z_2 - z_2 f \zb_1 &
\zb_1 f z_1 + z_2 f \zb_2 
\end{pmatrix} .
}
Now acting on the state $\bra{0,0}$, we have
\EQ{
\bra{0,0} U U\dag=  \begin{pmatrix}
0&0&0&0 \\ \cdots & \cdots & \cdots& \cdots\\
 \cdots & \cdots & \cdots& \cdots\\ \cdots & \cdots & \cdots& \cdots
\end{pmatrix} .
}
Obviously the completeness relation is not satisfied!
Notice that 
an ansatz of the form  \eq{udeff}, where there is an overall factor
of the shift operator $u\dag$,  works well for the case of
$U(1)$. There is no reason to restrict oneself to this form\footnote{
In fact it is easy to find a zero mode of $P$ which cannot be written as
$\tilde U_0 (\b_p u\dag \ket{s}).$ This zero mode is given by
$U_{\rm new}\begin{pmatrix}0\\ \ket{0,0}\end{pmatrix},$ where $U_{\rm new}$
refers to the right hand side of Eq.~\eqref{Ucor}.
} 
for higher
$U(N)$. Indeed a more general solution which does not take this form 
can be written down,
\EQ{ \label{Ucor}
U =  \begin{pmatrix}
0 & \sqrt{\frac{z_1\zb_1 +z_2\zb_2 + \z}{z_1\zb_1 +z_2\zb_2 + \r^2 +\z} }\\
(z_1\zb_1 +z_2\zb_2 )\b_p u\dag & 0 \\
-\sqrt{\z + \r^2} \zb_2 \b_p u\dag & \r z_1 
\frac{1}{\sqrt{(z_1\zb_1 +z_2\zb_2+ \z)(z_1\zb_1 +z_2\zb_2 +\r^2 + \z)}} \\
-\sqrt{\z + \r^2} \zb_1 \b_p u\dag & -\r z_2 
\frac{1}{\sqrt{(z_1\zb_1 +z_2\zb_2+ \z)(z_1\zb_1 +z_2\zb_2 +\r^2 + \z)}}
\end{pmatrix} \ .
} 
A special feature of this solution is that the shift
operator $u\dag$ appears only in the first column of $U$, where $\b_p$
appears.  It is not difficult to show that this is indeed a general
feature for the form of $U$ in the nonabelian case. 

It is easy to check that \eq{Ucor} satisfies
\EQ{
\bar \Delta U =0, \quad  \bar U U = 1_{\sst [2]\times [2]}.
}
Moreover one can check that the ADHM completeness relation
\eqref{cmpl}
\EQ{ 
1-U\bar U=\Delta f\bar\Delta
}
is satisfied,  and the field-strength
$F_{mn}$ is anti-selfdual.

We further note that at large distances $z_i\zb_i \gg\zeta $, 
\eq{Ucor} coincides
with the matrix $U$ of the commutative instanton \eqref{ueq}.
It follows that at distances large compared to noncommutativity scale,
the noncommutative instanton gauge field coincides with the commutative
instanton in the singular gauge \eqref{sutin},\eqref{singin}.
On the other hand, at short
distances the noncommutativity parameter $\zeta$ regulates the short-distance
singularity in \eqref{singin}.
Hence we get a regular noncommutative ASD instanton
which at large distances looks like the BPST instanton in the singular gauge.
This means that the LSZ reduction formulae can be applied as usual
to the functional integral representation
of the instanton Green functions
for calculating instanton contributions to  effective actions
(see e.g. \cite{MO1}).

Finally we calculate the topological charge $Q$. Using \eq{canform}
and \eq{asdu},  it is easy to obtain
\EQ{\label{trf2}
{\rm Tr}_N F^2 = -16 {\rm Tr}(A^2 +D^2 + 4 A D -2 B C) f^2,
} 
where we have denoted
\EQ{
\bar b U \bar U b =
\begin{pmatrix}
(\r^2+ \z)\zb_2 \b_p^2 z_2 
+ \r^2 z_1 f g\zb_1 &
(\r^2+ \z)\zb_2 \b_p^2 z_1 
- \r^2 z_1 fg \zb_2 \\
 (\r^2+ \z)\zb_1 \b_p^2 z_2 
- \r^2 z_2 fg \zb_1 &
(\r^2+ \z)\zb_1 \b_p^2 z_1 
+ \r^2 z_2 fg \zb_2
\end{pmatrix}
:= 
\begin{pmatrix}
A & B \\
C & D
\end{pmatrix}
}
and 
\EQ{
g= \frac{1}{z_1 \zb_1 +z_2 \zb_2 + \z} .
}
Substituting \eq{trf2} into the definition \eq{topc}, we find
\SP{
Q=-\sum_{N=1}^\infty
& {4\,(N+1)\over N(N+2{a}^{2}+2)^{2}(N+2{a}^{2}+1)^{2}
 (N+2 ) (N+2{a}^{2}+3)^{2}(N+2{a}^{2})} \\
&\times(748{a}^{4}{N}^{2}
+144{a}^{2}N+350{a}^{2}{N}^{2}+332{a}^{4}N
+230{a}^{2}{N}^{3}+ 62{a}^{2}{N}^{4}
+ 432{a}^{6}N+272{a}^{8}N\\
&+716\,{a}^{6}{N}^{2}+304{a}^{8}{N}^{2}
+406{a}^{4}{N}^{3}+280{a}^{6}{N}^{3}+84{a}^{4}{N}^{4}
+48{a}^{10}{N}^{2} \\
&+72{a}^{8}{N}^{3}+64{a}^{10}N
+6{a}^{2}{N}^{5}+6{a}^{4}{N}^{5}+36{a}^{6}{N}^{4}+36N-296{a}^{6}
-160{a}^{8}\\
&-32{a}^{10}+37{N}^{3}
-72{a}^{2}-240{a}^{4}+60{N}^{2}
+{N}^{5}+10{N}^{4}) \, = \, -1 \ , \label{qres2}
}
where $a=\rho/\sqrt{\zeta},$ and $N=n_1+n_2.$ Note that the
dependence on the instanton modulus $\rho$ disappears in the final answer.
Here we disagree with the results of \cite{KLY}--versions 1-3 which reported
$\rho$-dependence in $Q$ for the case at hand.
Our result \eqref{qres2}, as expected,
is in agreement with general argument of \S\ref{sec:S23} that
$|Q|=k.$

\subsection{SD instanton in the SD background on 
$\bf{R}_{\rm NC}^4$}\label{sec:S53}

Here we investigate the SD $1$-instanton solution of the $U(N)$
noncommutative theory in the SD-$\th$ background. This solution
has been previously studied in \cite{Furuuchi3}.

Without  loss of generality, we specialize here to the minimal case of
gauge group $U(2)$.
The matrix $\Delta$ is determined by solving the unmodified ADHM constraints
\eqref{sdiasdt} and reads:

\EQ{\Delta=\begin{pmatrix}
\rho & 0 \\ 0&\rho \\
z_2 & z_1 \\
-\zb_1 & \zb_2 \end{pmatrix} \ , \qquad
\bar\Delta=\begin{pmatrix}
\rho & 0 &  \zb_2 & -z_1 \\
0 & \rho & \zb_1 & z_2
 \end{pmatrix} \ .
\label{del1sd}} The factorization relation follows, and \EQ{f=
{1\over z_1\zb_1+z_2\zb_2+\rho^2+\zeta/2} \ . \label{fdefsd}} The
normalized matrix $U$ can now be constructed directly: \EQ{U=
\begin{pmatrix} -\zb_2\sqrt{1\over z_1\zb_1 +z_2\zb_2 +\rho^2 }
& \,&
z_1\sqrt{1\over z_1\zb_1 +z_2\zb_2 +\zeta +\rho^2 } \\
-\zb_1 \sqrt{1\over z_1\zb_1 +z_2\zb_2 +\rho^2 } &\, &
-z_2\sqrt{1\over z_1\zb_1 +z_2\zb_2 +\zeta +\rho^2 } \\
\rho \sqrt{1\over z_1\zb_1 +z_2\zb_2 +\rho^2 } &\, & 0\\
0 &\, & \rho \sqrt{1\over z_1\zb_1 +z_2\zb_2 +\zeta +\rho^2 }
 \end{pmatrix} \ , \quad
\bar\Delta U=0 \ , \quad \bar U U=1 \ .
\label{usdsd}
}
Note that in the SD/SD case at hand there are no states annihilated by
$U$ and no shifts operators are introduced in \eqref{usdsd}
in distinction with the ASD/SD cases considered earlier.

One can check with a straightforward calculation
that the ADHM completeness relation
\eqref{cmpl}
is satisfied, $1-U\bar U=\Delta f\bar\Delta$; the field-strength
$F_{mn}$ is selfdual, giving rise to an instanton number $Q=1$
by the Corrigan's identity.

Now we want to investigate the behaviour of the SD instanton at
large distances, $z_i\zb_i \gg\zeta,$
where noncommutativity can be neglected.
It follows from \eqref{usdsd} that in this limit
the noncommutative instanton coincides with the commutative
BPST instanton in the {\it regular} gauge,
\EQ{A_n= \bar U \partial_n U\, \rightarrow
\ {i  \ \bar\eta^a_{nm}\ x^m \ \tau^a
\over  x^2 + \rho^2}
\ . \label{bpstlim}
}
It is interesting to compare this SD/SD instanton with the ASD/SD
solution discussed earlier. While the former approaches the regular-gauge BPST
instanton, the latter tends to the singular-gauge BPST anti-instanton.
One might wonder if it is possible to gauge-transform the SD/SD instanton
in such a way that it tends to the singular-gauge BPST SD instanton.
In the commutative set-up one can always pass from the regular-gauge 
SD instanton
to the
singular gauge SD instanton with a singular $SU(2)$ gauge transformation,
$S^\dagger=x_m\sigma^m /\sqrt{x^2}.$
The noncommutative generalization of $S^\dagger$ is
\EQ{S^\dagger=\begin{pmatrix}
z_2 & z_1 \\ -\zb_1 & \zb_2\end{pmatrix}
{1\over \sqrt{z_1\zb_1 +z_2\zb_2 +\zeta/2}}
\ . }
It is easy to see that
this is not a unitary operator,
\EQ{S S^\dagger=1\ ,
\quad S^\dagger S= \begin{pmatrix}
1-|0,0\rangle \langle 0,0|&\, & 0 \\
0&\,& 1  \end{pmatrix} \ ,
}
hence $S^\dagger$ is not an allowed gauge transformation on the Hilbert space
$\cH$.

The LSZ reduction formulae cannot be applied directly to the
gauge field component $A_n$ of the SD/SD instanton, since it does not
fall off sufficiently fast at large distances.
However the LSZ amputation rules can still be applied to the
field strength and to the scalar-field and fermion-field
components of the instanton supermultiplet.
This is all what is required in e.g. deriving instanton contributions
to the Seiberg-Witten prepotential in the commutative \cite{MO1}
and the noncommutative case \cite{HKT}.

\subsection{Instanton  on
$\bf{R}_{\rm NC}^2\times \bf{R}^2$ }\label{sec:S531}

Finally we present the normalized ADHM matrix $U$
for the case of $\bf{R}_{\rm NC}^2\times \bf{R}^2$ 
(cf. \eqref{Ucor}):
\EQ{ \label{Ucrr}
U =  \begin{pmatrix}
0 & \sqrt{\frac{z_1\zb_1 +z_2\zb_2 + \z}{z_1\zb_1 +z_2\zb_2 + \r^2 +\z} }\\
(z_1\zb_1 +z_2\zb_2 )\b  & 0 \\
-\sqrt{\z + \r^2} \zb_2 \b  & \r z_1 
\frac{1}{\sqrt{(z_1\zb_1 +z_2\zb_2+ \z)(z_1\zb_1 +z_2\zb_2 +\r^2 + \z)}} \\
-\sqrt{\z + \r^2} \zb_1 \b  & -\r z_2 
\frac{1}{\sqrt{(z_1\zb_1 +z_2\zb_2+ \z)(z_1\zb_1 +z_2\zb_2 +\r^2 + \z)}}
\end{pmatrix} \, 
\begin{pmatrix} g^\dagger & 0\\ 0 & 1 \end{pmatrix} \,
 \ .
} 
which leads to a regular instanton.
Here $g^\dagger$ is the same singular $U(1)$ gauge transformation
as in \eqref{gdes} and the normalization factor $\beta$ is
(cf. \eqref{betdn}):
\EQ{\beta= 
 1_{\sst [2]\times [2]}{1\over
\sqrt{(z_1\zb_1 +z_2\zb_2)(z_1\zb_1 +z_2\zb_2+\zeta+\rho^2)}} \ .
\label{nbetdn}}
Again the topological charge is equal to minus one as a consequence of 
\eqref{qcalc}.

\section*{Acknowledgements}
We thank Tim Hollowood, Hugh Osborn,
Stefan Theisen and Tatsuya Ueno 
for comments and useful discussions.
We also thank Hyun Seok Yang for discussions on the earlier version
of this paper and on \cite{KLY}, \cite{K2}.
This work was partially supported by a PPARC SPG grant.

\end{document}